\documentclass[aps,pra,longbibliography,twocolumn,superscriptaddress,floatfix,notitlepage]{revtex4-2}
\usepackage[T1]{fontenc}
\usepackage{graphicx}
\usepackage{bbm}
\usepackage{float} 
\usepackage{color}
\usepackage[usenames,dvipsnames]{xcolor}
\usepackage{amsmath} 
\usepackage{amstext}
\usepackage{latexsym}
\usepackage[colorlinks=true,citecolor=NavyBlue,linkcolor=RubineRed,urlcolor=NavyBlue]{hyperref}
\usepackage{lipsum}
\usepackage{enumitem}
\usepackage{graphicx}

\usepackage{amssymb}

\usepackage{amsfonts}
\usepackage{epsfig}
\usepackage{dsfont}
\usepackage{mathrsfs}

\newcommand{\beq}{\begin{equation}}
\newcommand{\eeq}{\end{equation}}
\newcommand{\beqa}{\begin{eqnarray}}
\newcommand{\eeqa}{\end{eqnarray}}
\newcommand{\normord}[1]{:\mathrel{#1}:}
\usepackage[utf8]{inputenc}

\def\fakeH{\rlap{\hspace{-0.05cm}\'{}\hspace{-0.1cm}\'{}}}
\begin{document}

\title{Exact Dynamics of the 
Tomonaga-Luttinger Liquid and Shortcuts to Adiabaticity}

\author{L\'eonce Dupays   \href{https://orcid.org/0000-0002-3450-1861}{\includegraphics[scale=0.05]{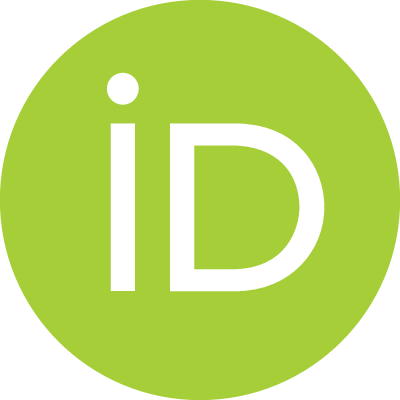}}}
\email{leonce.dupays@gmail.com}
\affiliation{Department  of  Physics  and  Materials  Science,  University  of  Luxembourg,  L-1511  Luxembourg, Luxembourg}

\author{Bal\'azs D\'ora  \href{https://orcid.org/0000-0002-5984-3761}{\includegraphics[scale=0.05]{orcidid.eps}}}
\affiliation{Department of Theoretical Physics, Institute of Physics, Budapest University of Technology and Economics, M\fakeH{u}egyetem rkp. 3, H-1111 Budapest, Hungary}
\author{Adolfo del Campo  \href{https://orcid.org/0000-0003-2219-2851}{\includegraphics[scale=0.05]{orcidid.eps}}}
\email{adolfo.delcampo@uni.lu}
\affiliation{Department  of  Physics  and  Materials  Science,  University  of  Luxembourg,  L-1511  Luxembourg, Luxembourg}
\affiliation{Donostia International Physics Center,  E-20018 San Sebasti\'an, Spain}

\begin{abstract}
Controlling many-body quantum systems is a highly challenging task required to advance quantum technologies. Here, we report progress in controlling gapless many-body quantum systems described by the Tomonaga-Luttinger liquid (TLL). To do so, we investigate the exact dynamics of the TLL induced by an arbitrary interaction quench, making use of the $SU(1,1)$ dynamical symmetry group and the Schr\"odinger picture. First, we demonstrate that this approach is useful to perform a shortcut to adiabaticity, that cancels the final non-adiabatic residual energy of the driven TLL and is experimentally implementable in the semiclassical limit of the sine-Gordon model. Second, we apply this framework to analyze various driving schemes in finite time, including linear ramps and smooth protocols. 
\end{abstract}
\maketitle
\section{Introduction}
In one spatial dimension, the Fermi liquid theory breaks down and collective effects appear \cite{ Haldane1981}. In this scenario, the Tomonaga-Luttinger liquid (TLL) \cite{Tomonaga1950, Luttinger1963} provides a universal low-energy description of different quantum many-body systems. As such, it makes possible to describe rapardigmatic models such as the Lieb-Liniger model \cite{LiebLiniger1963}, the Tonks-Girardeau gas \cite{Tonks1936, Girardeau1960}, the Fermi gas \cite{Bacsi2020}, and the Bose-Hubbard model in the continuum \cite{Cazalilla2003b}.  The TLL facilitates the description of interacting fermions and bosons in terms of bosonic collective degrees of freedom \cite{Mattis1965}. 
Due to its broad applicability, the TLL has served as a test-bed for studying nonequilibrium quantum phenomena in both theoretical and experimental studies \cite{Cazalilla2011, Karrasch12, GuanBatchelor13,Bernier14, Cedergren2017,Yang2017, Vranje2018,Haller2020}.

The nonequilibrium dynamics of a TLL can be induced by an interaction quench.  The dynamics of a quenched TLL has shed light onto the long-time behavior of an integrable gas of hard-core bosons, its description by a generalized Gibbs ensemble, and the thermalization of isolated integrable systems \cite{Cazalilla2006, Cazalilla2016, Kinoshita2006, Rigol2007}. Quenches in the TLL have also been applied to the study of optimal control \cite{Rahmani11,Rahmani13}, time-ordered and out-of-time-order correlators, Loschmidt echoes, and work distributions \cite{Dora2017, Dora2014, Dora2013, Bacsi2013, Langmann17, Dora2020, Bacsi2019, Pollmann2013, Kennes2013, Nessi2013, nessi2015, Vu2020, Wei2021, Datta23}, motivating experimental efforts to test theoretical predictions \cite{experimentLuttinger2008, Rueg2008}. Another application is the engineering of many-particle quantum heat engines using a TLL as a working substance  \cite{ChenWatanabe2019,Gluza21}.  In classical thermodynamics, the maximum efficiency of a heat engine is independent of the working medium \cite{callen1985thermodynamics}. This need not be the case in the quantum realm, as demonstrated in the study of interaction-driven quantum cycles using a one-dimensional interacting Bose gas as a working medium \cite{ChenWatanabe2019}. An enhanced efficiency was predicted for a weakly interacting engine at the border of the quantum critical regime \cite{Yang2017}. However, the finite-time dynamics and thermodynamics of such interaction-driven engines is yet to be explored \cite{ChenWatanabe2019}. Collective nonequilibrium phenomena in quantum heat engines can lead to quantum supremacy \cite{Jaramillo16}, and their engineering can circumvent the trade-off between efficiency and power \cite{Beau16,delCampo18}.

Interaction quenches implemented in a finite time generally lead to the breakdown of adiabatic dynamics and generate residual excitations in the final state that are responsible for quantum friction and are often referred to as (dissipated) heat \cite{delcampo14,Jaramillo16,Deng18Sci}. In many applications, it is desirable to find fast-driving schemes that do not generate such excitations. In the study of the driven TLL,  the following questions naturally arise:
\begin{itemize}
\item What is the exact dynamics of the TLL following an interaction change in finite time? 
\item  Are there nonadiabatic protocols for driving the TLL that do not generate residual excitations in the final state?
\end{itemize}
These two questions are interwoven as knowledge of the exact dynamics may allow one to perform its reverse engineering, implementing a prescribed evolution, e.g., for quantum state preparation \cite{Chen10,delcampo11}. 

 Nonadiabatic driving protocols that avoid excitations in the final state without the requirement for slow driving are known as Shortcuts To Adiabaticity (STA) \cite{Chen10,Torrontegui13,GueryOdelin19}. 
For systems of continuous variables, the study 
and the use of STA in many-body systems has been greatly advanced in processes exhibiting scale invariance. This dynamical symmetry allows us to relate the density profile at two different times by a simple scaling transformation.  This symmetry also governs the evolution of any other local correlation function.
Its use is of fundamental relevance in probing of ultracold gases \cite{CastinDum96,Kagan96,Dalfovo99,Giorgini08,Gritsev10}. STA for quantum fluids have been developed exploiting this symmetry \cite{delcampo11,delcampo12,delcampo13,Deffner14,Papoular15,Jaramillo16,delcampo2021probing}, with theoretical proposals being accompanied by implementations in the laboratory \cite{Schaff10,Schaff11,Schaff11njp,Rohringer2015,Deng18,Deng18Sci,Diao18}. However, scale invariance is restricted to potentials accounting for the interatomic interactions with the same scaling dimension as the kinetic energy operator and is thus generally broken in many-body systems.  
In particular, it is broken in the one-dimensional Bose gas with finite interactions, a paradigmatic system for the experimental study of integrable systems and their nonequilibrium dynamics \cite{Rohringer2015}.  Likewise, in the TLL, scale invariance is preserved at the level of each mode, but it is broken for macroscopic observables.

In this work, we first provide a general framework to describe the exact dynamics of a  TLL in an arbitrary driving protocol. We then use this framework to engineer STA  protocols in the TLL without relying on scale invariance. Our work thus presents a significant advance in the understanding of TLL dynamics and the engineering of STA protocols beyond scale-invariant dynamics \cite{delcampo13,Deffner14} and without relying on perturbative methods.
 
Previous works describing the TLL dynamics often rely on the Heisenberg picture and are focused on perturbative solutions and sudden quenches \cite{Dora2011, Dora2013, Cazalilla2016}. Some studies have reported exact analytical solutions in specific scenarios, e.g., for a linear quench of the $g_{2}$ parameter, describing the scattering between left and right movers in the TLL \cite{Dora2011,Dora2014,Cazalilla2016}. However, such specific solutions lack versatility and neglect the scattering between right (resp. left)  movers with themselves, with coupling $g_{4}$.  

The Schr\"odinger picture provides an alternative equivalent representation of the dynamics that is particularly useful in the presence of invariants of motion \cite{Lewis1969,Lohe2008,Chen10,delcampo12,delcampo13,Deffner14}. Using it, we describe the exact dynamics of the TLL in an arbitrary driving scheme. Building on such a framework, we characterize the time evolution in novel driving schemes beyond the sudden-quench limit. In particular, we report frictionless STA control protocols with no residual excitations in the final state. We further describe the exact dynamics in the case $g_{2}=g_{4}$, which is experimentally relevant to ultracold interacting bosons or the XXZ spin chain. 

\section{Summary of methods and results}
Our work involves two conceptual and technical steps.
First, we introduce the driven TLL and its connection to the Lewis-Riesenfeld theory of invariants of motion and use this formalism to determine an STA in an interaction quench. This fast, nonadiabatic scheme avoids inducing excitations in the final state when the TLL dynamics is assisted by an auxiliary Hamiltonian term, obtained by tailoring the sine-Gordon potential in the semiclassical approximation. The proposed STA protocols rely on an exact solution for the dynamics of the semiclassical sine-Gordon model. Previous solutions of the time-dependent semiclassical sine-Gordon model involve a time-dependent Gaussian variational Ansatz \cite{Lovas2022}. Our solution considers the case in which both the sine-Gordon potential and the interaction strength are varied simultaneously and may find applications beyond STA. Its interest also lies in its simplicity, as the modulation of the Luttinger parameter is given by the solution to a harmonic oscillator equation where the frequency of the oscillator is proportional to the sine-Gordon gap. We present a protocol that allows us to interpolate the dynamics between two TLLs with different initial and final values of the sound velocity $v_{s}$ and the Luttinger parameter $K$. This protocol requires the sudden opening of a semiclassical sine-Gordon gap.  An alternative, smoother protocol is then provided to drive the TLL from its ground state to the ground state of another TLL. This scheme is valid for systems verifying translational Galilean invariance, such as the Heinsenberg-Ising chain and a variety of ultracold atomic systems. The analytical description of the dynamics we report is also useful to determine an ``accidental STA protocol'' \cite{Jaramillo16}, i.e., a STA obtained by a forward solution of the dynamics that is stopped at a given time of evolution to prepare the final stationary state, that is an eigenstate of the instantaneous Hamiltonian. This accidental STA  can be used to reduce the Luttinger parameter $K$, at variance with the other protocols presented, which are restricted to an increase of the Luttinger parameter to ensure the stability of the underlying semiclassical sine-Gordon Hamiltonian. We also show that the semiclassical sine-Gordon model arising in this context provides an effective field-theory description of the driven long-range Lieb-Liniger model with 1D Coulomb interactions.

Second, we demonstrate that the TLL description using invariants of motion makes it possible to describe exactly quantum quenches in the TLL. In particular, we show that determining the dynamics of the TLL for equal forward scattering is equivalent to solving the well-known Ermakov equation \cite{Lewis1969}, with a frequency given by the spectrum of the time-dependent TLL. Reducing the difficulty of solving the dynamics of the TLL to solving the Ermakov equation greatly eases the description of the driven-TLL time evolution and opens the door for finding many beautiful solutions for its dynamics. In particular, we obtain a new analytical solution for the Ermakov equation when the time-dependent frequency evolves linearly in time. Using it, we analyze in detail driving protocols from the noninteracting to the interacting regime and vice versa. We investigate the non-adiabatic residual energy resulting from an interaction potential that is linear in time, as well as from a smooth interpolating polynomial ramp. We further analyze a perturbative solution for small quenches and characterize its breakdown,  highlighting the relevance of the exact framework we introduce. We demonstrate a plateau behavior for short time quenches and a decay in $\ln(\tau_{Q})/\tau^{2}_{Q}$, where $\tau_{Q}$ is the duration of the quench, in the adiabatic limit for a linear quench. We also provide a lower bound on the adiabatic time for the TLL. Finally, we investigate the non-adiabatic residual energy for a reverse-engineering protocol where the modulation of the interaction is an inverse polynomial.

The paper is organized as follows. In Sec. \ref{Tomonaga_TLL}, we review the TLL and introduce notations. Section \ref{secLR} introduces the formalism of invariants of motion and its link with the TLL. In Sec. \ref{eq:section_shortcuts}, we study STA in the TLL assisted by the semiclassical sine-Gordon potential. In Sec. \ref{ermakov_equation_section}, we demonstrate the link between the Ermakov equation and the TLL. In Sec. \ref{section_linear}, we solve the exact dynamics for a linear quench and compare it with a perturbative approach that is restricted to linear quenches of small amplitude. In Sec. \ref{section_polynomial}, we study the dynamics of a smooth polynomial quench. In Sec. \ref{section_polynomial_quench}, we study another solution for the dynamics, for an interaction quench taking the form of an inverse polynomial. In Sec. \ref{section_summary}, we give a summary of our results and discuss further directions of research.

\section{The Time-Dependent TLL and Its $SU(1,1)$ Representation \label{Tomonaga_TLL}}
Let us consider the Hamiltonian of the TLL \cite{Tomonaga1950,Luttinger1963,Haldane1981}, the derivation of which is reviewed in App. \ref{app:TLL}. 
Its explicit form reads
\begin{align}
\label{eq:TomonagaLuttinger}
H_{\rm TL}(t)&=\sum_{p\neq 0}\hbar\omega(p,t)K_{0}(p)+\sum_{p\neq0}\frac{\hbar g(p,t)}{2}\left[K_{-}(p)+K_{+}(p)\right]\nonumber\\
&+C_0(t)-\sum_{p\neq 0}\frac{\hbar \omega(p,t)}{2},
\end{align}
where $v_{F}$ is the Fermi velocity and $g_{2/4}(p,t)$ denote the interaction potentials, so that
\begin{eqnarray}
\omega(p,t)&=& |p|\left[v_{F}+\frac{g_{4}(p,t)}{2\pi}\right],\\
g(p,t)&=& |p|\frac{g_{2}(p,t)}{2\pi}.
\end{eqnarray}
The TLL Hamiltonian (\ref{eq:TomonagaLuttinger}) is expressed in terms of the generators of the $\mathfrak{su}(1,1)$ algebra \cite{zhang2022quantum} 
\begin{eqnarray}
K_{0}(p)&=&\frac{1}{2}\left[b^{\dagger}(p)b(p)+b^{\dagger}(-p)b(-p)+\mathbbm{1}\right],\\
K_{+}(p)&=&b^{\dagger}(p)b^{\dagger}(-p), \quad K_{-}(p)=b(p)b(-p), 
\end{eqnarray}
that verify the commutation relations $[K_{0}(p),K_{\pm}(p)]=\pm K_{\pm}(p)$, $[K_{-}(p),K_{+}(p)]=2K_{0}(p)$, with the bosonic ladder operators $[b(p),b^{\dagger}(p')]=\delta_{pp'}$, and the zero momentum energy term $C_0(t)=\frac{\hbar g_{2}(0,t)}{4L}\left(N^{2}-J^{2}\right)+\frac{ \hbar g_{4}(0,t)}{4L}\left(N^{2}+J^{2}\right)+\frac{\pi \hbar v_{F}}{L}(N^{2}+J^{2})$. $N$ is the total number operator and $J$ the current operator. The interacting TLL (\ref{eq:TomonagaLuttinger}) can be diagonalized  through a Bogoliubov transformation 
\begin{align}U(\beta_{t})&=\exp\left\{\sum_{p\neq 0} \frac{\beta_{t}}{2}[K_{+}(p)-K_{-}(p)]\right\},\end{align} 
satisfying the condition $\tanh(2\beta_{t})=-\frac{g(p,t)}{\omega(p,t)}$.
We next define the time-dependent bosonic operators $a(p,t)=U(\beta_{t})b(p)U^{\dagger}(\beta_{t})$, where
\begin{align}
a(p,t)&=\cosh(\beta_{t})b(p)-\sinh(\beta_{t})b^{\dagger}(-p),\\
b(p)&=\cosh(\beta_{t})a(p,t)+\sinh(\beta_{t})a^{\dagger}(-p,t).
\end{align}
After the Bogoliubov transformation, the Hamiltonian (\ref{eq:TomonagaLuttinger}) takes the diagonal form  
\begin{eqnarray}
\label{eq:diagonal_Tomonaga_Luttinger}
H_{\rm TL}(t)=\sum_{p\neq0}\epsilon(p,t)a^{\dagger}(p,t)a(p,t)+E_{0}(t)+C_0(t),\nonumber
\end{eqnarray}
with $\epsilon(p,t)=\hbar\sqrt{[\omega(p,t)]^{2}-[g(p,t)]^{2}}$ and the ground state energy shift $E_{0}(t)=\frac{1}{2}\sum_{p\neq0}\left[\epsilon(p,t)-{\color{black}\hbar \omega(p,t)|p|}\right]$. As noticed in Ref. \cite{Haldane1981}, the energy shift induced by the quench does not diverge if $\epsilon(p,t)$ tends to $v_{F}|p|$ for high momenta, which is equivalent to assume an ultraviolet cutoff. The excited states of the bosonic basis can be constructed as tensor products of the eigenstate modes \cite{Haldane1981}
\begin{eqnarray}
\label{eq:eigenstates}
|\{n_{p}\},t\rangle=\bigotimes_{p\neq0}\frac{\left[a^{\dagger}(p,t)\right]^{n_{p}}}{\sqrt{n_{p}!}}|\Omega\rangle. \label{eq:eigenstate_TLL}
\end{eqnarray}
 From this analysis, it follows that modifying the interactions between the particles alters the spectrum and the eigenstates of the Hamiltonian, where $n_{p}$ is the occupation number of the $p$ mode, and $|\Omega\rangle$ the ground state of the noninteracting TLL. In particular, as the previous diagonalization is performed at a given time $t$, Eq. (\ref{eq:eigenstate_TLL}) corresponds to the instantaneous eigenstate of the interacting TLL Hamiltonian. Following the bosonization procedure \cite{Haldane1981}, the TLL can be decomposed according to a phase operator $\phi(x)$ and its conjugated density fluctuation operator $\Pi(x)=\frac{\partial_{x}\theta(x)}{\pi}$, verifying the canonical bosonization commutation relation $[\phi(x),\frac{\partial_{x'}\theta(x')}{\pi}]=i\delta(x-x')$ \cite{Haldane1981}. The meaning of these fields is directly seen in the expression of the field creation operator, given in the fermionic case by \cite{Cazalilla2004}
\begin{eqnarray}
\Psi^{\dagger}(x)\approx\sqrt{\rho_{0}-\frac{\partial_{x}\theta(x)}{\pi}}\sum_{p}e^{i(2p+1)(\pi\rho_{0}x-\theta(x))}e^{-i\phi(x)} \label{eq:bosonized_field},\nonumber\\
\end{eqnarray}
where the average density $\rho_{0}$ is related to the Fermi momentum by $\pi\rho_{0}=k_{F}$. There exists a direct mapping between the field operators and the bosonic basis as further explained in App. \ref{app:TLL},  
\begin{eqnarray}
&&\resizebox{.96\hsize}{!}{
$\theta(x)=-N\frac{\pi x}{L}-i\sum_{k\neq 0}\sqrt{\frac{\pi |k|}{2L}}\frac{e^{-\nu |k|/2-i k x}}{k}\left[b^{\dagger}(k)+b(-k)\right],$}\nonumber\\
\label{eq:theta_expression}\\
&&\resizebox{.95\hsize}{!}{$\phi(x)=J\frac{\pi x}{L}+i\sum_{k\neq 0}\sqrt{\frac{\pi |k|}{2 L }}\frac{e^{-\nu |k|/2-i k x}}{|k|}\left[b^{\dagger}(k)-b(-k)\right],$}\nonumber\\
\label{eq:phi_expression}
\end{eqnarray}
where the cutoff parameter $\nu$ ensures convergence  in  the limit $\nu \to 0$ \cite{Haldane1981}.  This momentum cutoff also provides a range of validity for the linear spectrum approximation so that $\frac{1}{\nu}\sim \pi \rho_{0}$ \cite{Buchner2003}.
It is worth considering a momentum-independent interaction potential satisfying $g_{2/4}(p,t)=\tilde{g}_{2/4}(t)$, where the tilde is used to clearly distinguish the coupling coefficients from $g_{2/4}(p,t)$. In such case, one can express the value of the velocity $v_{s}(t)$ and the Luttinger parameter $K(t)$ in terms of the interaction couplings in Eq. (\ref{eq:TomonagaLuttinger}) as
\begin{eqnarray}
K(t)&=&\sqrt{\frac{2\pi v_{F}+\tilde{g}_{4}(t)-\tilde{g}_{2}(t)}{2\pi v_{F}+\tilde{g}_{4}(t)+\tilde{g}_{2}(t)}},\\
v_{s}(t)&=&\sqrt{\left[v_{F}+\frac{\tilde{g}_{4}(t)}{2\pi}\right]^{2}-\left[\frac{\tilde{g}_{2}(t)}{2\pi}\right]^{2}}.
\end{eqnarray}
As a consequence, the TLL Hamiltonian can be equivalently written in terms of the fields $\phi(x)$ and $\theta(x)$ as
\begin{eqnarray}
\resizebox{.96\hsize}{!}{$
H_{\rm TL}(t)=\frac{\hbar v_{s}(t)}{2\pi}\int_{0}^{L}dx \left[K(t)(\partial_{x}\phi(x))^{2}+K^{-1}(t)(\partial_{x}\theta(x))^{2}\right].$}\label{eq:field_TLL}\nonumber
\end{eqnarray}
For a free fermionic Hamiltonian, $K=1$ and $v_{s}=v_{F}$. In the particular case when $\tilde{g}_{2}(t)=\tilde{g}_{4}(t)=\tilde{g}_{2,4}(t)$, one finds that $K^{-1}(t)=\sqrt{1+\frac{\tilde{g}_{2,4}(t)}{\pi v_{F}}}$ and $v_{s}(t)=v_{F}K^{-1}(t)$.  We note that the condition $|\tilde{g}_{2}(t)|<2\pi v_{F}+\tilde{g}_{4}(t)$ ensures that $v_{s}(t)$  remains positive \cite{schulz1998fermi}. The choice $\tilde{g}_{2}(t)=\tilde{g}_{4}(t)=\tilde{g}_{2,4}(t)$ is relevant in the context of ultracold atomic gases, as further described in App.~\ref{app:implementation_TLL}. It is also interesting to notice that the condition $v_{s}(t)=v_{F}K^{-1}(t)$ is valid for systems that verify translational Galilean invariance as demonstrated in Ref. \cite{Cazalilla2003b}. The condition $\tilde{g}_{2}(t)=\tilde{g}_{4}(t)$ also holds approximately for the Heisenberg-Ising chain \cite{Fowler_1980,schulz1998fermi} in the
weak coupling limit, though it breaks down with increasing Ising interaction in the exact solution of the model \cite{Giamarchibook}. More details about this spin-based implementation are discussed in App.~\ref{app:implementation_TLL}.

\section{The Driven TLL and Lewis-Riesenfeld Invariants \label{secLR}}
Back in 1969, in the study of the driven quantum oscillator, Lewis and Riesenfeld \cite{Lewis1969} recognized the importance of invariants of motion for the exact description of quantum dynamics.  They demonstrated that the knowledge of the spectral decomposition of an invariant of motion is enough to solve the time-dependent Schr\"odinger equation. When the time evolution is generated by a Hamiltonian $\mathcal{H}$, an operator $I(t)$ is an invariant of motion provided it satisfies  
\begin{eqnarray}
\label{eq:LiouvilleVonNeumann}
i\hbar \frac{d I(t)}{dt}=i\hbar\frac{\partial I(t)}{\partial t}-[\mathcal{H},I(t)]=0.\label{eq:invariant_equation}
\end{eqnarray}
 Making use of the spectral decomposition $I(t)=\sum_n\lambda_n|\lambda_{n},t\rangle\langle \lambda_{n},t|$ yields the solution to the time-dependent Schr\"odinger equation  in the form \cite{Lewis1969,Lohe2008}
\begin{eqnarray}
\label{eq:eigenstate_evolution}
|\Psi_{t}\rangle=\exp\left(i\int_{0}^{t} \kappa_{n}(s) ds\right)|\lambda_{n},t\rangle,
\end{eqnarray}
with the Lewis-Riesenfeld phase $\kappa_{n}(s)=\langle \lambda_{n},s |i \frac{\partial }{\partial s}-\frac{\mathcal{H}}{\hbar}|\lambda_{n},s\rangle$.
However, finding an invariant of motion for a given Hamiltonian can be difficult, particularly in many-body systems \cite{delcampo12,delcampo13,Deffner14,delcampo2021probing}. 

For quadratic Hamiltonians, such as the TLL, it is always possible to build the invariant of motion as a product of linear invariants, which is itself an invariant.
For instance, for the TLL, choosing $b(p,t)=f(p,t)b(p)+h^{*}(p,t)b^{\dagger}(-p)$ as a linear invariant, one can derive the quadratic invariant $I(t)=b^{\dagger}(p,t)b(p,t)$. Inserting the linear invariant in (\ref{eq:LiouvilleVonNeumann}) yields an equation of the form
\begin{eqnarray}
i\hbar\frac{\partial}{\partial t}\begin{pmatrix}f(p,t)\\h^{*}(p,t)\end{pmatrix}=&\mathcal{M}_{t}\begin{pmatrix}f(p,t)\\h^{*}(p,t)\end{pmatrix},
\end{eqnarray}
where 
\begin{eqnarray}
\mathcal{M}_{t}=\hbar\begin{pmatrix}-\omega(p,t)&g(p,t)\\-g(p,t)&\omega(p,t)
\end{pmatrix}.
\end{eqnarray}
To determine the exact dynamics, one can diagonalize this system of time-dependent differential equations. Let us write the diagonal matrix $\mathcal{D}_{t}=\mathcal{P}_{t}\mathcal{M}_{t}\mathcal{P}^{-1}_{t}$,  where $\mathcal{P}_{t}$ is a $2\times 2$ invertible matrix. It then follows that
\begin{eqnarray}
i\hbar\frac{\partial}{\partial t}\mathcal{P}_{t}\begin{pmatrix}f\\h^{*}\end{pmatrix}&=&\left(\mathcal{P}_{t}\mathcal{M}_{t}\mathcal{P}^{-1}_{t}+i\hbar\frac{\partial \mathcal{P}_{t}}{\partial t}\mathcal{P}^{-1}_{t}\right)\mathcal{P}_{t}\begin{pmatrix}f\\h^{*}\end{pmatrix}.\nonumber\\
\label{eq:dynamics_first_order}
\end{eqnarray}
Due to the time-dependence of the matrix $\mathcal{M}_{t}$, this system is, in general, not exactly solvable.
This approach is however useful in the adiabatic limit ($\frac{d \mathcal{P}_{t}}{dt}\mathcal{P}^{-1}_{t}\rightarrow 0$), or for a sudden quench, when $\mathcal{M}_{t}$ is constant. Knowledge of the spectrum of the quadratic invariant can then be directly deduced from the Bogoliubov transformation, leading to the solution of the Schr\"odinger equation. The system of coupled first-order differential equations  (\ref{eq:dynamics_first_order}) is similar to the one encountered for the equations of motion in the Heisenberg picture, which were numerically studied for the TLL in \cite{Bukov2012}, and for which several analytical solutions are given in \cite{Chudzinski2016}.

We present an alternative approach, circumventing the need to solve (\ref{eq:dynamics_first_order}),  which consists in describing the dynamics using an invariant of motion and several unitary transformations. Specifically, given a time-independent operator $I_0$ with a known spectral decomposition and a  unitary time-evolution operator $Q$, the operator  $I=QI_{0}Q^{\dagger}$ is an invariant of motion when Eq. (\ref{eq:invariant_equation}) is satisfied for a Hamiltonian of the form   
\begin{eqnarray}
\label{eq:invariant_Hamiltonian}
\mathcal{H}(t)=F(I,t)+i\hbar\frac{\partial Q}{\partial t}Q^{\dagger},
\end{eqnarray}
where $F(I,t)=I/\sigma^{2}_{t}$  and $\sigma_{t}$ a time-dependent scalar function. The dynamical phase can be expressed by direct substitution of (\ref{eq:eigenstate_evolution}) into the Schr\"odinger equation with Hamiltonian (\ref{eq:invariant_Hamiltonian}), from which one obtains the identity $\kappa_{n}(s)=-\frac{1}{\hbar}\langle \lambda_{n},s|F(I,s)|\lambda_{n},s\rangle=-\frac{1}{\hbar}\langle \lambda_{n},0|I_{0}|\lambda_{n},0\rangle/\sigma^{2}_{s}$.  We build on the analogy with the time-dependent harmonic oscillator \cite{Lohe2008} and quantum fluids with $SU(1,1)$ dynamical symmetry group, manifested in scale invariance, i.e., self-similar dynamics \cite{Sutherland98,Gritsev10,delcampo11,delcampo12,delcampo13,Deffner14,delcampo16,Jaramillo16,Beau16,Beau20,delcampo2021probing}. Consider the choice $I_{0}=\sum_{p \neq 0}\hbar\omega_{0p}K_{0}(p)$, 
together with the unitary time-evolution operator $Q=V(\zeta_{p})U(\gamma_{p})$, where 
\begin{eqnarray}
U(\gamma_{p})&=&\exp\left\{\sum_{p\neq 0}\frac{\ln(\gamma_{p})}{2}[K_{+}(p)-K_{-}(p)]\right\}, \label{eq:U(rho)}\\
V(\zeta_{p})&=&\exp\left\{\sum_{p\neq 0}\frac{i\zeta_{p}}{2\omega_{0p}}[2K_{0}(p)+K_{+}(p)+K_{-}(p)]\right\},\label{eq:V(alpha)}\nonumber\\
\end{eqnarray}
so that $I=QI_{0}Q^{\dagger}$. As shown in the App.~\ref{app:calculations_TDTL}, one obtains from Eq.  (\ref{eq:invariant_Hamiltonian}) the explicit form of the time-dependent Hamiltonian
\small
\begin{align}
\label{eq:main_equation}
\mathcal{H}(t)
&=Q\left(\sum_{p\neq 0}\frac{K_{0}(p)\hbar\omega_{0p}}{\sigma^{2}_{t}}\right)Q^{\dagger}(\zeta_{p})+i\hbar\frac{\partial Q}{\partial t}Q^{\dagger}\nonumber\\
&=\sum_{p\neq0}\hbar \{K_{0}(p)+\frac{1}{2}[K_{+}(p)+K_{-}(p)]\}\nonumber\\
&\times\Big\{\frac{\omega_{0p}}{2}\frac{1}{(\sigma_{t}\gamma_{p})^{2}}+\frac{ 2\zeta^{2}_{p}}{\omega_{0p}}\left(\frac{\gamma_{p}}{\sigma_{t}}\right)^{2}-\frac{\dot{\zeta_{p}}}{\omega_{0p}}-\frac{2\zeta_{p}}{\omega_{0p}}\frac{\dot{\gamma}_{p}}{\gamma_{p}}\Big\}\nonumber\\
&+\sum_{p\neq 0}\hbar \{K_{0}(p)-\frac{1}{2}[K_{+}(p)+K_{-}(p)]\}\frac{\omega_{0p}}{2}\left(\frac{\gamma_{p}}{\sigma_{t}}\right)^{2}\nonumber\\
&+i\hbar\sum_{p\neq 0}\left[K_{+}(p)-K_{-}(p)\right]\left\{\frac{\dot{\gamma}_{p}}{2\gamma_{p}}-\zeta_{p}\left(\frac{\gamma_{p}}{\sigma_{t}}\right)^{2}\right\}.
\end{align}\normalsize
As a consequence, the time evolution of the eigenstates of Hamiltonian (\ref{eq:main_equation}) is given by
\begin{eqnarray}
|\{n_{p}\},t\rangle &=&e^{-\frac{i}{\hbar}\int_{0}^{t}\sum_{p\neq 0}\frac{\epsilon(p,0)}{\sigma^{2}_{s}}\left(n_{p}(0)+\frac{1}{2}\right)ds}Q(t)|\{n_{p}\},0\rangle.\nonumber\\
\label{eq:dynamical_state}
\end{eqnarray}
Here,  $n_{p}(0)$ denotes the occupation number, and $\epsilon(p,0)$ is the $p$-mode energy eigenvalue at the initial time.
The Hamiltonian (\ref{eq:main_equation}) does not exactly match the Hamiltonian of the TLL (\ref{eq:TomonagaLuttinger}) unless further assumptions on the coefficients $\gamma_{p}$, $\sigma_{t}$, and $\zeta_p$ are invoked. Yet, with an appropriate choice of these coefficients, this Hamiltonian provides the means to describe a wide variety of exact solutions of the driven TLL.  In particular, we will demonstrate that a specific choice of the coefficients provides an exact solution for the dynamics of the semi-classical sine-Gordon model. We will further see that an alternative choice of coefficients also leads to an exact solution for the dynamics of the TLL.
\section{Shortcuts to Adiabaticity in the TLL Assisted by a Semiclassical Sine-Gordon Potential \label{eq:section_shortcuts}}
STAs constitute a versatile set of techniques for the fast preparation of quantum states without the requirement for slow driving \cite{Torrontegui13,GueryOdelin19}. STAs have been designed for this purpose in interacting ultracold gases that exhibit scale-invariant dynamics, such as one-dimensional hard-core bosons, a two-dimensional Bose-Einstein condensate, and a three-dimensional unitary Fermi gas \cite{delcampo11,delcampo13,Deffner14,Papoular15,Beau16,Beau20,delcampo2021probing}. These theoretical insights have been accompanied by significant experimental progress \cite{Schaff10,Schaff11,Schaff11njp,Rohringer2015,Deng18,Deng18Sci,Diao18,Ness18}.  Finding STA for many-body quantum systems beyond scale-invariant dynamics remains a challenge, in particular, when avoiding approximate methods, such as the use of variational ansatze and mean-field theories \cite{Muga09,MasudaNakamura10,Torrontegui13,Masuda14}. 

A prominent instance in which scale-invariance is broken concerns ultracold Bose gases tightly confined in a waveguide. As shown by Olshanii \cite{Olshanii98}, such systems are accurately described by a mathematical model known as the Lieb-Liniger gas \cite{LiebLiniger1963}.
This integrable model describes a one-dimensional Bose gas with contact delta-type interactions that break scale-invariance in the time evolution for any nonzero finite value of the interaction strength \cite{Buljan08,Buljan08b,delcampo11epl,delcampo11,Iyer12,Iyer13,Rohringer2015}. 
Can STAs be engineered in this context? 

The TLL formalism provides an efficient low-energy description of the Lieb-Liniger gas. In addition, the formalism introduced in the previous section is useful in determining an STA protocol for interaction quenches in the TLL. Using it, we next engineer nonadiabatic STA protocols that drive the system from one eigenstate of the initial TLL to the respective eigenstate of a final TLL. To achieve this, the nonadiabatic evolution is assisted by an auxiliary driving term that takes the form of a semiclassical sine-Gordon potential, provided that the low-energy description still applies in the driven regime. 

\subsection{General exact solution $g_{2}(p,t)\neq g_{4}(p,t)$ of the semiclassical sine-Gordon model}
\begin{figure}[t]
\begin{centering}
\includegraphics[width=\linewidth]{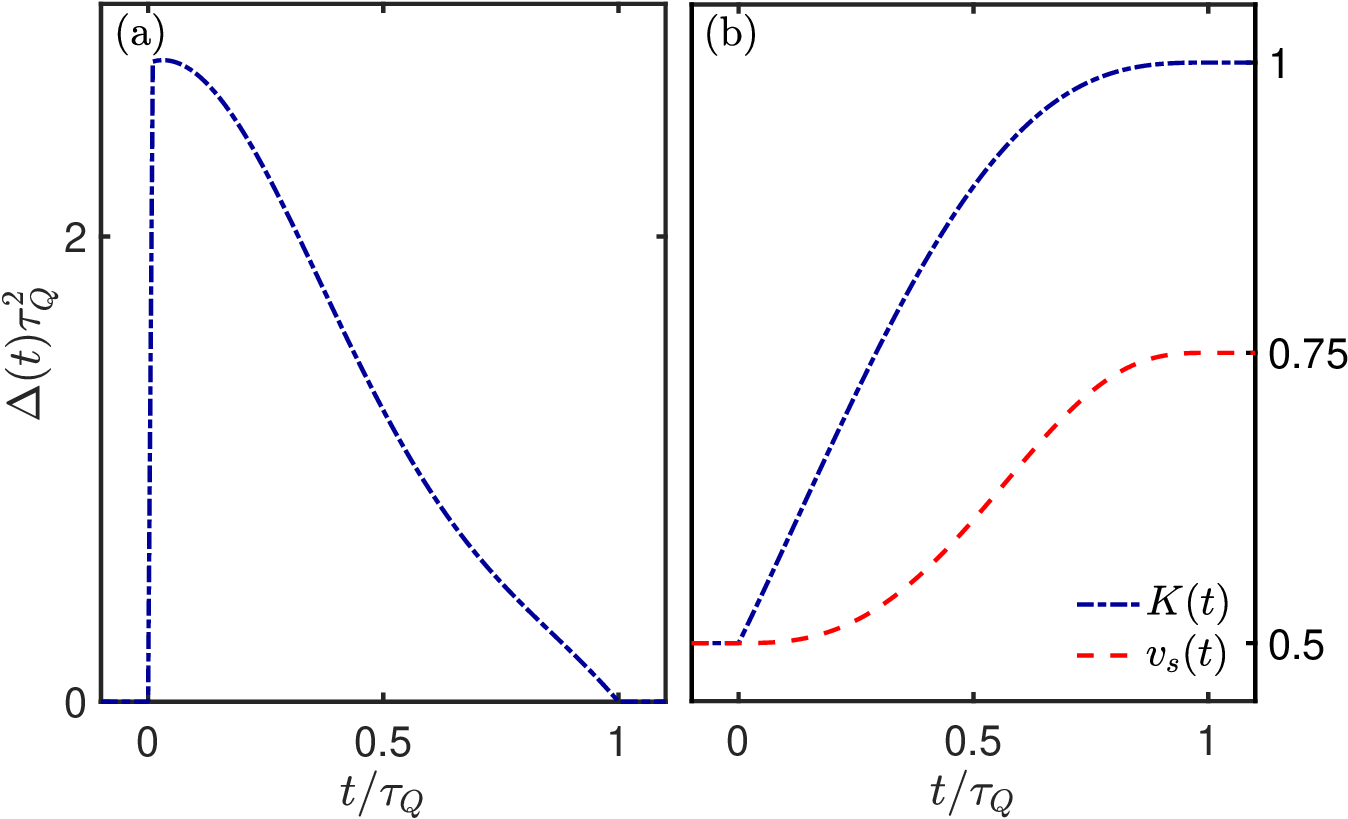}
\caption{Finite-time driving in the semiclassical sine-Gordon model from one TLL to another TLL with a suddenly quenched gap. Panel (a) illustrates the modulation of the sine-Gordon gap term in the semiclassical limit. Panel (b) depicts the associated evolution of the Luttinger coefficients during the quench. We chose the protocols $\gamma=\gamma_{0}+[\gamma_{f}-\gamma_{0}]\mathcal{P}_{4}(t/t_{f})$ and $\sigma_{t}=\sigma_{0}+[\sigma_{f}-\sigma_{0}]\mathcal{P}_{5}(t/t_{f})$ with $\sigma_{0}=1/\gamma_{0}=\sqrt{2}$ and $\gamma_{f}=\sigma_{f}\sqrt{3}/2=1$. And the fifth-order polynomial $\mathcal{P}_{5}(s)=10 s^3-15s^4+6s^5$ and the fourth-order polynomial $\mathcal{P}_{4}(s)=s^4-2s^3+2s$.}
\label{gapped_to_gapless}
\end{centering}
\end{figure}

Let us start by describing a general solution for the dynamics of the semiclassical sine-Gordon model \cite{Giamarchibook}. Choosing the parameters $\gamma_{p}=\gamma$ and $\sigma_{t}$ as scalar real functions and $\zeta_{p}=\frac{\dot{\gamma}\sigma^{2}_{t}}{2\gamma^{3}}$ provides exact dynamics for the semiclassical sine-Gordon model. The Hamiltonian can be directly expressed as a function of the fields $\phi(x)$ and $\theta(x)$ upon identifying $\omega_{0p}=v_{F}|p|$ and $K(t)=\gamma^{2}$ and $v_{s}(t)=v_{F}/\sigma^{2}_{t}$, in the form
\begin{eqnarray}
\label{eq:sine_gordon}
\mathcal{H}(t)&=&\frac{\hbar v_s(t)}{2}\int_{0}^{L}\left[\frac{\pi}{K(t)}\Pi(x)^{2}+\frac{K(t)}{\pi}(\partial_{x}\phi(x))^{2}\right]dx\nonumber\\
&+&\Delta(t)\frac{\hbar}{\pi v_{F}}\int_{0}^{L}dx[\theta(x)]^{2},\nonumber\\
\end{eqnarray}
with $\Pi(x)=\partial_{x}\theta(x)/\pi$ and 
\begin{equation}
\Delta(t)=\left(\frac{\sigma_{t}}{\gamma}\right)^{2}\left[\left(\frac{\dot{\gamma}}{\gamma}\right)^{2}-\frac{1}{2}\frac{\ddot{\gamma}}{\gamma}-\frac{\dot{\gamma}}{\gamma}\frac{\dot{\sigma}_{t}}{\sigma_{t}}\right]. 
\end{equation}
This Hamiltonian corresponds to the semiclassical approximation of the sine-Gordon model Hamiltonian \cite{Haldane1981PRL,Foini2017} with energy spectrum
\begin{eqnarray}
 \epsilon(p,t)&=&\hbar\frac{\gamma}{\sigma_{t}}\sqrt{\frac{\omega^{2}_{0p}}{(\sigma_{t}\gamma)^{2}}+2\Delta(t)}.\nonumber
\end{eqnarray}
In the context of quantum control for fast preparation of quantum states, the performance of a driving protocol can be benchmarked by quantifying the final nonadiabatic excitations.
The explicit evaluation of the nonadiabatic mean energy for the semiclassical approximation of the sine-Gordon model (\ref{eq:sine_gordon}) is described in App.~\ref{eq:mean_energy} for a pure state in the thermodynamic limit, and yields 
\footnotesize
\begin{align}
&\langle \mathcal{H}(t)\rangle\nonumber\frac{2\pi}{L}\\
&=\int_{0}^{+\infty}\left\{\frac{\hbar v_{F}|p|}{\sigma^{2}_{t}}-\frac{\hbar\sigma^{2}_{t}}{v_{F}|p|}\left[\frac{1}{2}\frac{\ddot{\gamma}}{\gamma}+\frac{\dot{\gamma}}{\gamma}\frac{\dot{\sigma}_{t}}{\sigma_{t}}-\frac{3}{2}\left(\frac{\dot{\gamma}}{\gamma}\right)^{2}\right]\right\}e^{-R_{0}p}dp,\nonumber\\
\end{align}\normalsize
where $R_{0}$ is introduced as a high energy cutoff. For slow protocols satisfying $\dot{\gamma}=\ddot{\gamma}=0$,  the dynamics converges to the adiabatic ramping of a TLL, described by the Bogoliubov transformation, in which the adiabatic energy for an initial pure state is given by 
\begin{align}
\langle\mathcal{H}(t)\rangle_{ad}=\frac{L}{2\pi}\frac{\hbar v_{s}(t)}{R^{2}_{0}}.
\end{align}
Driving the system from one TLL to another in equation (\ref{eq:sine_gordon}) is conveniently described by the cancellation of the boundary conditions $\dot{\gamma}(0)=\ddot{\gamma}(0)=\dot{\gamma}(\tau_{Q})=\ddot{\gamma}(\tau_{Q})=0$. For example, these boundary conditions can be engineered with a fifth-order polynomial ansatz for $\gamma$. However, one needs to carefully choose the driving protocol to ensure the positivity of the spectrum, which guarantees the stability of the semiclassical sine-Gordon model. As a consequence, the condition $\Delta(t)\geq 0$ has to be verified during the dynamics. To do so, we choose a driving protocol with a fourth-order polynomial for $\gamma$, which involves a sudden quench of $\Delta(t)$ at the initial time, followed by a smooth modulation. Fig.~\ref{gapped_to_gapless} provides an instance of a driving protocol from one eigenstate of a TLL to the respective eigenstate of another TLL.  The protocol starts with a sudden opening of the gap of the semiclassical sine-Gordon model and a subsequent gradual reduction of the gap according to the prescribed polynomial trajectory for $\gamma$ and $\sigma_{t}$. 
%
%
For this protocol, we consider that the system is initially prepared in a given eigenstate of the initial TLL, e.g., the ground state.  After the finite quench time $\tau_{Q}$, the mean energy of the Hamiltonian (\ref{eq:sine_gordon}) matches the mean energy of the adiabatically prepared state, given that $\ddot{\gamma}(\tau_{Q})=\dot{\gamma}(\tau_{Q})=0$, i.e., 
\begin{align}
\langle\mathcal{H}(\tau_{Q})\rangle=\frac{L}{2\pi}\frac{\hbar v_{s}(\tau_{Q})}{R^{2}_{0}}.
\end{align}
In other words, the driving protocol supplemented with the modulation of an external semi-classical sine-Gordon potential succeeds in canceling the nonadiabatic excitations in the final state, realizing an STA. As a result, the mean energy upon completion of the quench equals the adiabatic value, associated with an adiabatic ramping of the interactions in the TLL. To explore the implementation of this protocol in a specific experimental setting, we next focus on the case of equal interactions $g_{2}(p,t)=g_{4}(p,t)$, where the TLL provides a description for the low energy of the Lieb-Liniger model.
\subsection{Exact solution for $g_{2}(p,t)=g_{4}(p,t)$ of the semiclassical sine-Gordon model and experimental implementation in ultracold bosons \label{lattice_implementation}}
To make the protocol suitable for implementation in ultracold gases, it is convenient to focus on the case with  $\sigma^{2}_{t}=\gamma^{2}$ and $\omega_{0p}=v_{F}|p|$. The Hamiltonian (\ref{eq:sine_gordon}) then simplifies to
\begin{eqnarray}
\mathcal{H}(t)&=&\frac{\hbar v_s(t)}{2}\int_{0}^{L}\left[\frac{\pi}{K(t)}\Pi(x)^{2}+\frac{K(t)}{\pi}(\partial_{x}\phi(x))^{2}\right]dx\nonumber\\
&-&\frac{\hbar}{2\pi v_{F}}\left[\frac{1}{2}\frac{\ddot{K}(t)}{K(t)}-\frac{1}{4}\left(\frac{\dot{K}(t)}{K(t)}\right)^{2}\right]\int_{0}^{L}dx[\theta(x)]^{2}, \label{eq:sine_gordon_g2g4}\nonumber\\
\end{eqnarray}
where we recall $K(t)=\gamma^{2}$ and $v_{s}(t)=v_{F}(K(t))^{-1}$, as well as $\Delta(t)=-\ddot{\gamma}/(2\gamma)=-\frac{1}{4}\frac{\ddot{K}(t)}{K(t)}+\frac{1}{8}\left(\frac{\dot{K}(t)}{K(t)}\right)^{2}$. 

The Hamiltonian (\ref{eq:sine_gordon_g2g4}) can be approximately implemented in experiments with ultracold gases by inserting a lattice potential with time-dependent amplitude and a constant phase $V(x,t)=V_{0}(t)\cos(\lambda x+\phi_{0})$. The phase allows for the modulation of the sign of the potential, making it attractive or repulsive so that the supplementary potential term reads
\begin{eqnarray}
H_{V}&=&\int_{0}^{L}dx\rho(x)V(x,t).
\end{eqnarray}
The interaction Hamiltonian can be bosonized by using the harmonic fluid approximation for the density
\begin{eqnarray}
\resizebox{.95\hsize}{!}{$\rho(x)\approx\left(\rho_{0}-\frac{\partial_{x}\theta(x)}{\pi}\right)\left\{1+2\sum_{m=1}^{\infty}\cos[2m(\pi\rho_{0}x-\theta(x))]\right\}$}.\nonumber
\end{eqnarray}
Retaining only the terms that couple at first order, with the phase $\phi_{0}=\pi$, one obtains the semiclassical approximation of the sine-Gordon Hamiltonian, as further described in App.~\ref{app:implementation_TLL}
\begin{align}
H_{{\rm sG}}&=H_{\rm TL}+H_{V}\nonumber\\
&\approx\frac{\hbar v_{s}(t)}{2 \pi}\int_{0}^{L} dx \left[ K(t)(\partial_{x}\phi(x))^{2}+K^{-1}(t)\left(\partial_{x}\theta(x)\right)^{2}\right]\nonumber\\
&+2\hbar V_{0}(t)\rho_{0}\int_{0}^{L}dx[\theta(x)]^{2}.\label{eq:sg_model}
\end{align}
The quadratic term in $\theta(x)$ is justified from an expansion when  $V(x)$ is a deep optical lattice. This approximation is reminiscent of the use of lattice methods for describing the loading of optical lattices when starting from the continuum \cite{Masuda14,Sheikhan15,Zhou18}. Essentially, the fundamental system describing the ultracold bosons with time-dependent interactions in the lattice potential is then provided by the triplet of equations
\begin{align}
\ddot{\gamma}+4\pi v_{F} \rho_{0} V_{0}(t)\gamma&=0,\label{eq:differential_eq}\\
K(t)&=\gamma^{2},\\
v_{s}(t)&=v_{F}/\gamma^{2}.
\end{align}
These equations provide a direct relation between the time-dependent modulation of the potential $V_{0}(t)$ and the interaction strength embedded into the Luttinger parameter $K(t)$. As a consequence, if one chooses the modulation of the interactions by varying  $K(t)$ and $v_{s}(t)$,  a modulation of $V_{0}(t)$ is required. This protocol is tantamount to the reverse-engineering of the dynamics and determines the consistency conditions for realizing a prescribed quantum state trajectory. By contrast, if one chooses the modulation for $V_{0}(t)$, it is possible to solve the differential equation (\ref{eq:differential_eq}) for $\gamma$ to determine how to modulate $K(t)$ and $v_{s}(t)$.  This is a forward scheme in which the quantum state trajectory is not prescribed.
However, the possible dynamics are still restricted due to the stability of the semiclassical sine-Gordon model. The stability condition of the semiclassical sine-Gordon is given by the positivity of the spectrum. Diagonalizing the Hamiltonian in Eq. (\ref{eq:sine_gordon_g2g4}), one obtains the  spectrum 
%
\begin{figure}[t]
\begin{centering}
\includegraphics[width=\linewidth]{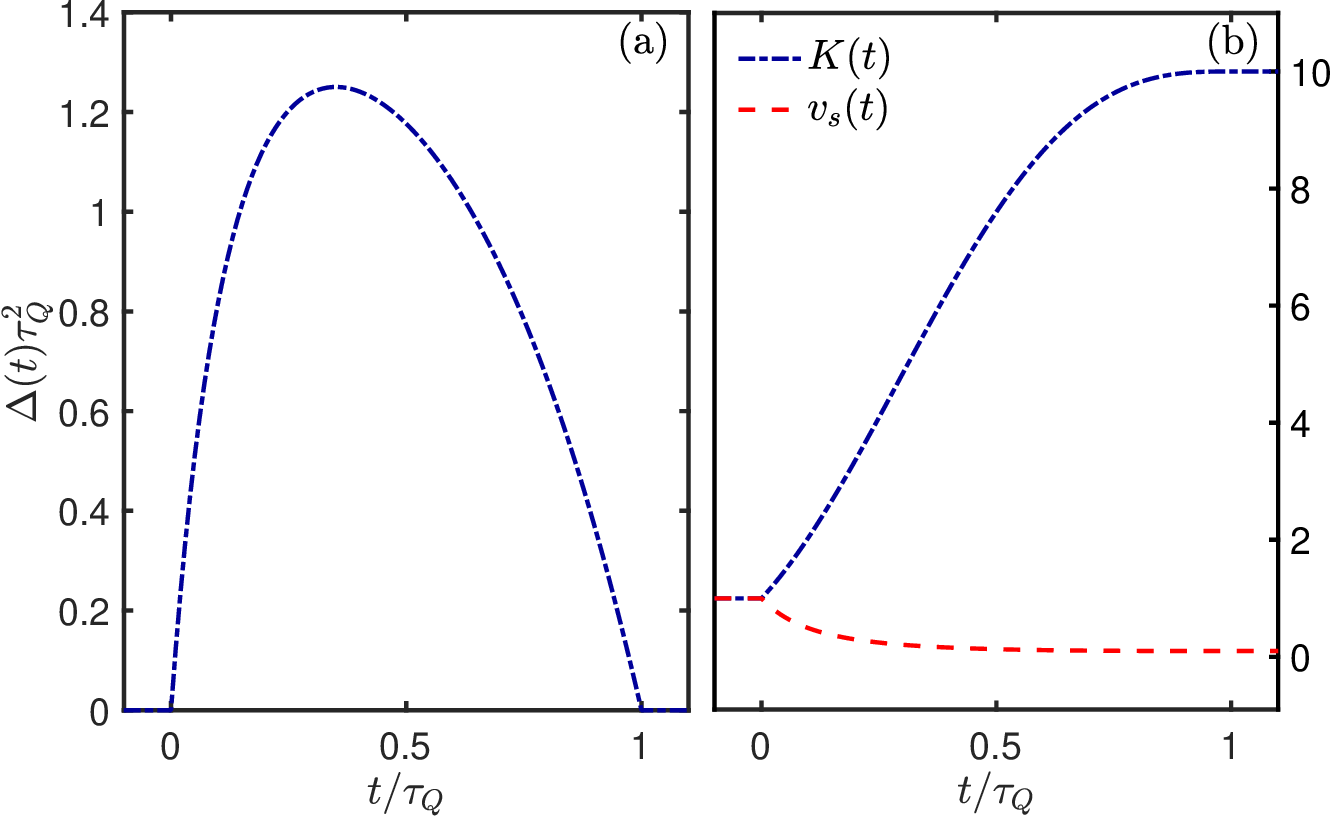}
\caption{Driving in the semiclassical sine-Gordon model between two gapless phases. Such transition is possible between two distinct TLLs with different Luttinger coefficients and without non-adiabatic residual energy. Panel (a) illustrates the modulation of the sine-Gordon gap term in the semiclassical limit. Panel (b) depicts the associated evolution of the Luttinger coefficients during the quench. We chose the protocol $\gamma=\gamma_{0}+[\gamma_{f}-\gamma_{0}]\mathcal{P}_{4}(t/t_{f})$ with $\gamma_{0}=1$ and $\gamma_{f}=\sqrt{10}$. And the fourth-order polynomial $\mathcal{P}_{4}(s)=s^4-2s^3+2s$, in particular we notice that $\ddot{\mathcal{P}}_{4}(0)=\ddot{\mathcal{P}}_{4}(1)=0$.
\label{gapless_to_gapless}}
\end{centering}
\end{figure}
\begin{eqnarray}
\epsilon(p,t)=\hbar\sqrt{p^{2}c^{*2}-{\rm sgn\left(\ddot{\gamma}/\gamma\right)}(m^{*}c^{*2})^{2}}, \label{epsilonpt}
\end{eqnarray}
where ${\rm sgn}(x)$ is the sign function.
This equation describes the spectrum of a massive relativistic particle with an effective speed of light $c^{*}=\frac{v_{F}}{\gamma^{2}}$ and mass $m^{*}=\frac{1}{c^{*2}}\sqrt{\frac{|\ddot{\gamma}|}{|\gamma|}}$. To ensure the stability of the TLL, the spectrum has to remain positive with $p^{2}c^{*2}>{\rm sgn(\ddot{\gamma}/\gamma)}(m^{*}c^{*2})^{2}$. In the thermodynamic limit, this condition can only be verified whenever $\ddot{\gamma}/\gamma<0$. In the forward dynamics protocol, the equation $-\frac{\ddot{\gamma}}{\gamma}=4\pi v_{F}\rho_{0}V_{0}(t)$ implies that if one chooses a positive potential the stability condition is always satisfied. However, in a reverse-engineering protocol, this is not necessarily guaranteed.
\begin{figure}
  \centering
  \includegraphics[width=1\columnwidth]{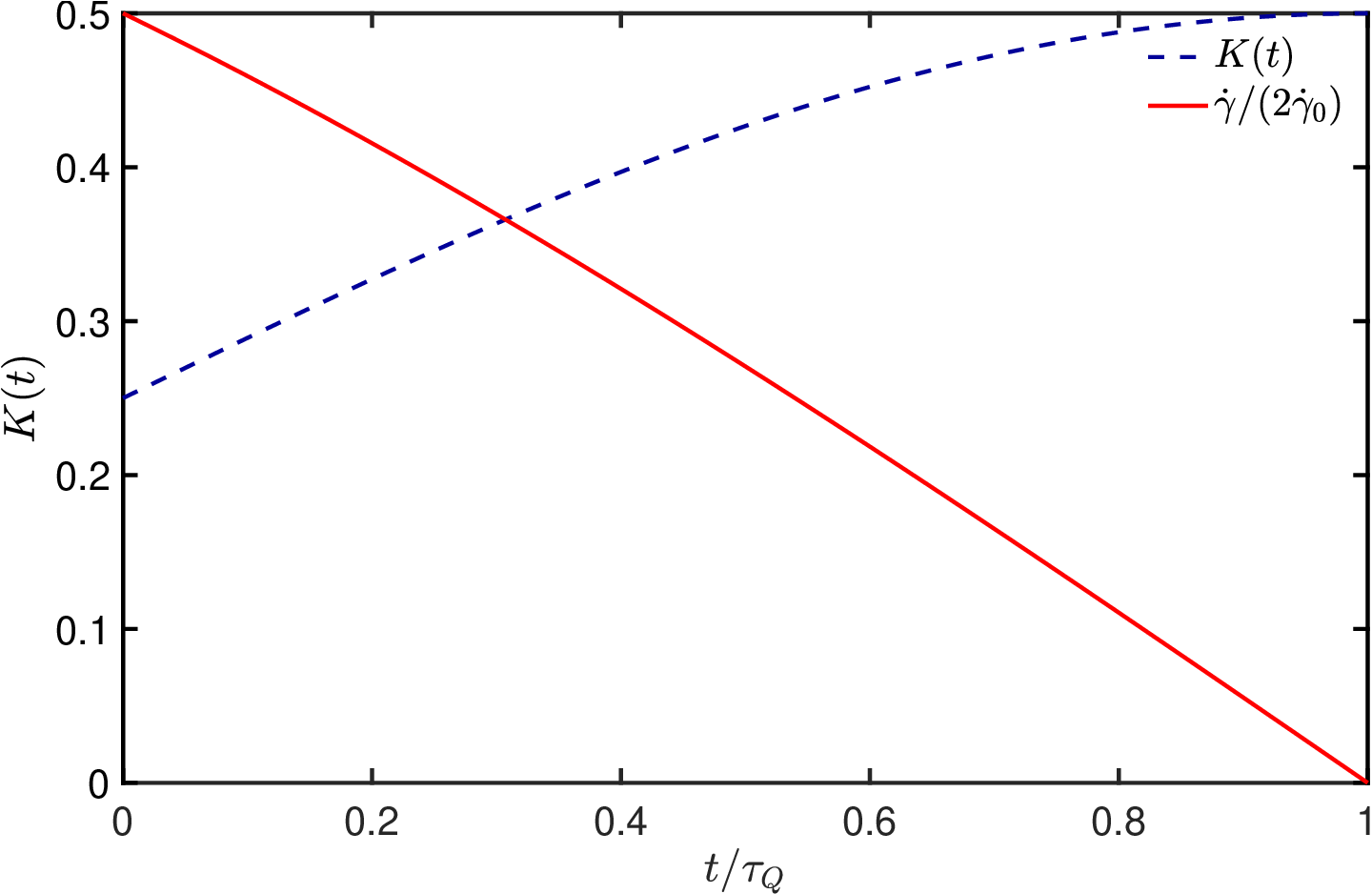}
  \caption{Modulation of the Luttinger parameter $K(t)$ for a constant sine-Gordon potential. We used the property $k_{F}=\pi \rho_{0}$ with $\hbar=m=1$  and conveniently choose the constant potential $V_{0}=10E_{F}$, with $E_{F}=\frac{k^{2}_{F}}{2m}$, $v_{F}=1$ and initial condition $\dot{\gamma}_{0}=10$ and $K(0)=1/4$. The Luttinger parameter is displayed as $K(t)=\gamma^{2}$ where $\gamma$ is given by Eq. (\ref{eq=solution_constant_potential}). It is interesting to look at the evolution of $\dot{\gamma}$. In particular, setting the duration of the protocol  $\tau_{Q}$ to the value $t_0$ in Eq. (\ref{tnsta}) leads to an accidental STA at the end time, with $\dot{\gamma}(\tau_Q)=0$.
  \label{fig:modulation_luttinger_constant}}
\end{figure}
\subsection{Reverse-engineering of the dynamics}
Let us first consider a reverse-engineering protocol. An interesting protocol is $\gamma(t)=\gamma_{0}+[\gamma_{f}-\gamma_{0}]\mathcal{P}_{4}(t/\tau_{Q})$ with $\mathcal{P}_{4}(s)=s^{4}-2s^{3}+2s$ that ensures $\ddot{\gamma}_{0}=\ddot{\gamma}_{f}=0$ and $\dot{\gamma}_{f}=0$. $\ddot{\gamma}/\gamma$ remains negative at all times if $\gamma_{f}>\gamma_{0}$. Fig.~\ref{gapless_to_gapless} illustrates the application of this protocol,  assisted by the semiclassical approximation of the sine-Gordon model, for driving a strongly interacting TLL to a TLL with weaker interactions. The nonadiabatic driving of the Luttinger coefficients is compensated by opening the gap during the dynamics, leading to a null non-adiabatic residual energy. 
\subsection{``Accidental'' Shortcuts to Adiabaticity}
Let us next consider the forward trajectory protocol described in solving the second-order differential equation (\ref{eq:differential_eq}). The example of a constant frequency is informative and leads to the well-known oscillatory solution 
\begin{eqnarray}
\gamma=\gamma_{0}\cos(4\pi v_{F}V_{0}\rho_{0}t)+\frac{\dot{\gamma}_{0}}{4\pi v_{F}V_{0}\rho_{0}}\sin(4\pi v_{F}V_{0}\rho_{0} t).\label{eq=solution_constant_potential}
\end{eqnarray}
There exist specific values of time $t_{n}$ when the derivative vanishes $\dot{\gamma}=0$, 
\begin{eqnarray}
\label{tnsta}
t_{n}=\frac{\arctan\left(\frac{\dot{\gamma_{0}}}{\gamma_{0}4\pi v_{F}V_{0}\rho_{0}}\right)+n \pi}{4\pi v_{F}V_{0}\rho_{0}}.
\end{eqnarray}
Terminating the driving protocol at any time $t_n$  provides an accidental STA protocol \cite{Beau16}, i.e., an STA obtained for a specific duration by a forward solution of the dynamics. Considering an initial TLL with Luttinger parameter $K(0)$, one can suddenly turn on a constant potential $V_{0}$ until the time $\tau_{Q}=t_{n=0}$, 
while simultaneously modulating the Luttinger parameter, according to the trajectory prescribed by the solution Eq. (\ref{eq=solution_constant_potential}). %
When the constant potential is turned off at the final time $t=\tau_Q$, the TLL is in a stationary state. Fig.~\ref{fig:modulation_luttinger_constant} provides an example of such protocol for the modulation of the Luttinger parameter $K(t)$ for a constant potential $V_{0}$, ensuring that at the end of the protocol, the derivative cancels $\dot{\gamma}=0$. Note that it is essential to take the initial value $\dot{\gamma}_{0}\neq 0$ so that the Luttinger parameter can evolve over one period. 

\subsection{Linear Ramping of the Lattice Potential}
Another protocol of interest is the linear ramp of the interaction potential $V_{0}(t)=\alpha t$, which is turned off to 
\begin{figure}
[H]
  \centering
\includegraphics[width=1\columnwidth]{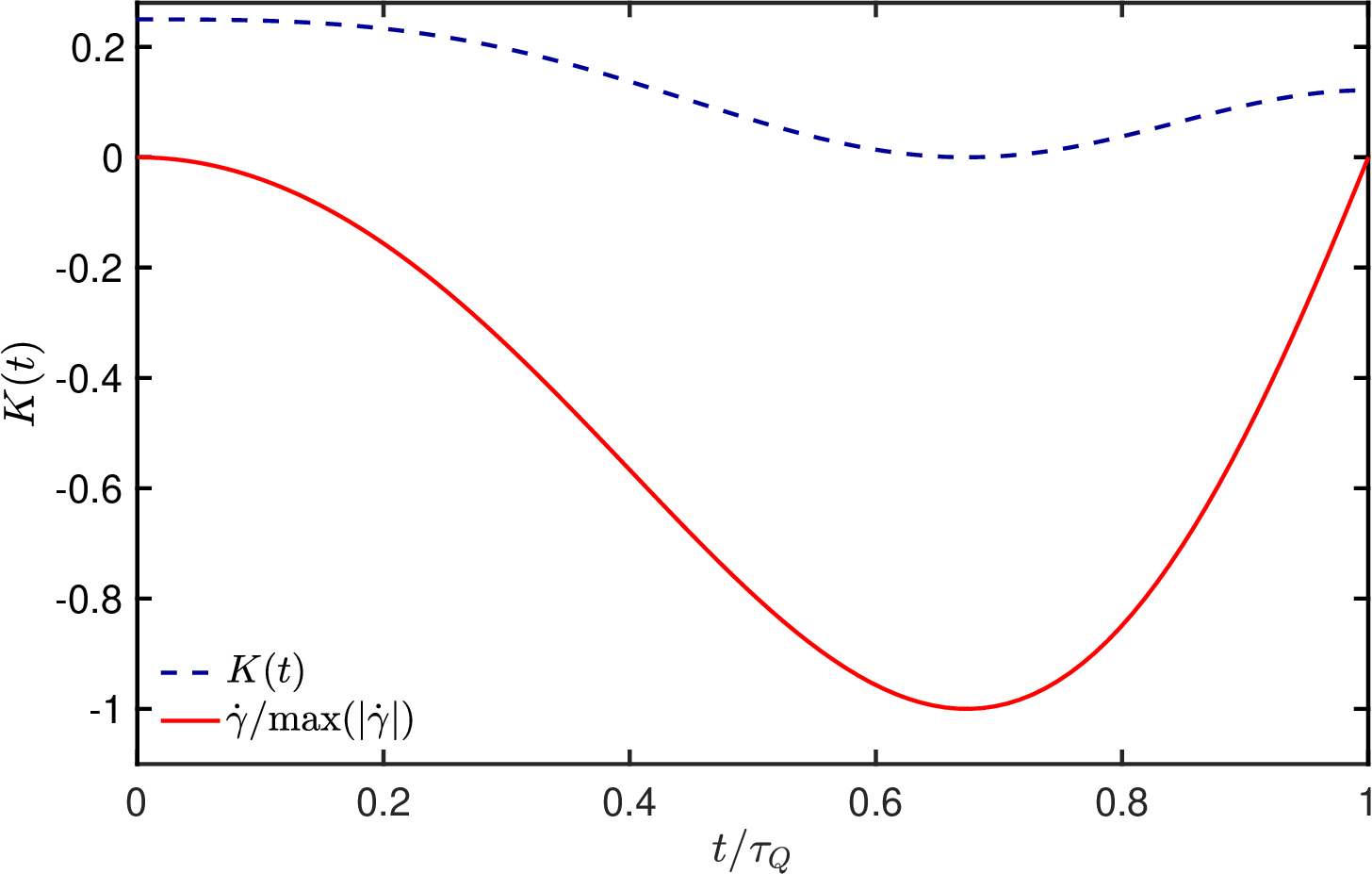}
  \caption{Modulation of the Luttinger parameter $K(t)$ for a linear ramp of the sine-Gordon potential during the quench time $\tau_{Q}$. At the end of the quench, the potential is set to zero. We used the property $k_{F}=\pi \rho_{0}$ with $\hbar=m=1$  and choose $d=4\pi v_{F}\alpha\rho_{0}$, $\alpha=3$, together  with the initial conditions $\dot{\gamma}_{0}=0$ and $K(0)=1/4$. The Luttinger parameter is displayed as $K(t)=\gamma^{2}$ where $\gamma$ is given by Eq. (\ref{eq=solution_linear_potential}). The evolution of $\dot{\gamma}$ governs the residual excitations, and the final time $\tau_{Q}$ is determined numerically as the root of $\dot{\gamma}(\tau_{Q})=0$. This accidental STA protocol reduces in finite-time the Luttinger parameter.
  \label{fig:modulation_luttinger_linear}}
\end{figure}
\noindent zero at the end of the quench time, whenever $\dot{\gamma}=0$. This leads to the Airy differential equation 
\begin{eqnarray}
\ddot{\gamma}+4\pi v_{F}\alpha\rho_{0}t\gamma=0,
\end{eqnarray}
with well-known solutions 
\small
\begin{align}
\gamma&=\pi \gamma_{0}\{\partial_{y}{\rm Bi}[y(0)]{\rm Ai}[y(t)]-\partial_{y}{\rm Ai}[y(0)]{\rm Bi}[y(t)]\},\label{eq=solution_linear_potential}\\
\dot{\gamma}&=\pi \gamma_{0}(-d)^{1/3}\{\partial_{y}{\rm Bi}[y(0)]\partial_{y}{\rm Ai}[y(t)]-\partial_{y}{\rm Ai}[y(0)]\partial_{y}{\rm Bi}[y(t)]\},\nonumber
\end{align}\normalsize
in terms of the Airy functions ${\rm Ai}(y)$ and ${\rm Bi}(y)$, with $ y(t)=(-d)^{1/3}t$, $d=4\pi v_{F}\alpha\rho_{0}$, and the choice $\dot{\gamma}_{0}=0$. This protocol has an advantage in that it allows for the reduction of the Luttinger parameter $ K(\tau_{Q})\leq K(0)$. By contrast, in the two accidental STA protocols proposed, the value of the second derivative is not null at the end of the quench $\ddot{\gamma}(\tau_{Q})\neq 0$, which may lead to nonzero residual energy in the final state. In this protocol, however, the potential is set to zero after the quench, and the final state becomes a stationary eigenstate of the final Hamiltonian as $\dot{\gamma}(\tau_{Q})=0$. Fig.~\ref{fig:modulation_luttinger_linear} provides an example of such protocol for the modulation of the Luttinger parameter $K(t)$ for a linear ramp potential $V_{0}(t)$, ensuring that at the end of the protocol the derivative cancels $\dot{\gamma}(\tau_{Q})=0$. The duration of the quench is chosen so that $\dot{\gamma}(\tau_{Q})=0$. 

\subsection{Simulation of the Long-Range Lieb-Liniger Model \label{long_range}}
It is interesting to note that the semiclassical sine-Gordon model of Eq. (\ref{eq:sine_gordon_g2g4}) can also be interpreted as a TLL with a long-range interaction. This is apparent in the momentum representation of (\ref{eq:sine_gordon_g2g4})
\begin{eqnarray}
\mathcal{H}(t)&=\frac{\hbar v_{F}}{2\gamma^{4}}\sum_{p\neq0}|p|[K_{0}(p)+\frac{1}{2}(K_{+}(p)+K_{-}(p))]\nonumber\\
&+\frac{\hbar v_{F}}{2}\sum_{p\neq 0}|p|[K_{0}(p)-\frac{1}{2}(K_{+}(p)+K_{-}(p))]\nonumber\\
&-\frac{\hbar \ddot{\gamma}}{v_{F}2\gamma}\sum_{p\neq0}\frac{1}{|p|}[K_{0}(p)+\frac{1}{2}(K_{+}(p)+K_{-}(p))],\nonumber\label{eq:semiclassical_SG}
\end{eqnarray} 
which can be identified as the TLL in Eq. (\ref{eq:TomonagaLuttinger}) with an interaction potential  of the form
\small
\begin{eqnarray}
\label{eq:solution}
\frac{g_{2}(p,t)}{2\pi}=\frac{g_{4}(p,t)}{2\pi}=\frac{g_{2,4}(p,t)}{2\pi}&=&\frac{\tilde{g}_{2,4}(t)}{2\pi}-\frac{1}{2v_{F}|p|^{2}}\left(\frac{\ddot{\gamma}}{\gamma}\right),\nonumber\\
\end{eqnarray}\normalsize
with $\frac{\tilde{g}_{2,4}(t)}{2\pi}=\frac{v_{F}}{2}\left(\frac{1}{\gamma^{4}}-1\right)$.
In the coordinate representation, the interaction potential reads
\begin{eqnarray}
\label{LRLLint}
\frac{g_{2,4}(x-x',t)}{2\pi}&=&\frac{\tilde{g}_{2,4}(t)}{2\pi}\delta(x-x')+\frac{|x-x'|}{8v_{F}}\left(\frac{\ddot{\gamma}}{\gamma}\right).\nonumber\\
\end{eqnarray}
Thus, the interaction (\ref{LRLLint}) combines a contact potential with the 1D Coulomb interaction. The 1D Coulomb gas is largely studied in statistical physics \cite{Lenard1963,Demery2012,Demery_2016}, where the experimental implementation is proposed by the use of charged infinite condenser sheets that can freely cross each other \cite{Lenard1963}, called the lattice fluid capacitor \cite{Demery2012,Demery_2016}. The 1D electric field can be engineered following ref. \cite{keldysh} by surrounding the 1D system with medium possessing a much smaller dielectric constant than that of the 1D tube. This is precisely the form of the recently introduced long-range Lieb-Liniger model \cite{delCampo2020PRR,Beau20prl,Yang2022}. The TLL model can describe fermionic or bosonic gases \cite{Haldane1981,Haldane1981PRL,Monien1998}. For fermions, $s$-wave scattering is forbidden by symmetry,
leading to the absence of this contact potential. 
However, for a Bose gas, the combination of short and long-range interactions has been proposed \cite{delCampo2020PRR,Beau20prl,Yang2022}. The long-range linear interaction is also reminiscent of meson confinement observed in condensed matter physics \cite{Kormos2017}.

The expression (\ref{LRLLint}) also clarifies why $\ddot{\gamma}/\gamma<0$ is required to stabilize the system. For positive $\ddot{\gamma}/\gamma$, the interaction favors keeping the particles together and results in phase separation, which is reminiscent of the ferromagnetic region of the Heisenberg-Ising chain with zero magnetization. In this region, bosonization breaks down, which is signaled by the imaginary $\epsilon(p,t)$ in Eq. \eqref{epsilonpt}.
On the other hand, for negative $\ddot{\gamma}/\gamma$, this interaction keeps the particles apart and opens up a gap in the spectrum, as in the antiferromagnetic region of the Heisenberg-Ising chain.  
\section{The Driven TLL and the Ermakov Equation \label{ermakov_equation_section}}
In the previous section, we made the assumption that the parameter $\gamma$ is momentum independent. However, lifting this assumption and considering a general momentum dependence for the parameter $\gamma_{p}$ leads to an exact solution for the TLL. In this section, we determine the exact dynamics for the driven TLL by solving a momentum-dependent Ermakov equation. Choosing the parameters $\sigma^{2}_{t}=\gamma^{2}_{p}$ and $\zeta_{p}=\frac{1}{2}\dot{\gamma}_{p}/\gamma_{p}$ in (\ref{eq:main_equation}) leads to
\begin{eqnarray}
\label{eq:dynamical_hamiltonian}
\mathcal{H}(t)&=&\sum_{p\neq 0}\hbar\omega(p,t)K_{0}(p)+\sum_{p\neq 0}\frac{\hbar g(p,t)}{2}[K_{+}(p)+K_{-}(p)]\nonumber\\
\end{eqnarray}
with $g(p,t)=\frac{\omega_{0p}}{2\gamma^{4}_{p}}-\frac{1}{2\omega_{0p}}\left(\frac{\ddot{\gamma}_{p}}{\gamma_{p}}\right)-\frac{\omega_{0p}}{2}$ and $\omega(p,t)=\frac{\omega_{0p}}{2\gamma^{4}_{p}}-\frac{1}{2\omega_{0p}}\left(\frac{\ddot{\gamma}_{p}}{\gamma_{p}}\right)+\frac{\omega_{0p}}{2}=\omega_{0p}+g(p,t)$.  One can now map the exact solution obtained from the invariants of motion to the TLL Hamiltonian. Identifying  (\ref{eq:dynamical_hamiltonian}) and (\ref{eq:TomonagaLuttinger}),
 with the choice $\omega_{0p}=v_{F}|p|$ determines the interaction strength
\begin{eqnarray}
g(p,t)&=&\frac{g_{2}(p,t)}{2\pi}|p|=\frac{g_{4}(p,t)}{2\pi}|p|\\
&=&\left\{-\frac{v_{F}}{2}+\frac{v_{F}}{2 \gamma^{4}_{p}}-\frac{1}{2v_{F}|p|^{2}}\left(\frac{\ddot{\gamma}_{p}}{\gamma_{p}}\right)\right\}|p|.
\end{eqnarray}
 Note that the TLL is recovered under the assumption $g_{2}(p,t)=g_{4}(p,t)$. In addition, the momentum dependence of $\gamma_{p}$ is in stark contrast with the momentum-independent case discussed in the previous section. As a result, it is now possible to obtain a solution for $\gamma_{p}$ so that $g(p,t)$ is linear in $|p|$, leading to a delta interaction potential in position space. To completely recover the TLL Hamiltonian, one needs to take into account the energy shift $\mathcal{H}(t)=H_{\rm TL}(t){\color{black} +\sum_{p\neq 0}\frac{\hbar \omega(p,t)}{2}}$. As in the time-dependent harmonic oscillator, determining the exact dynamics of the TLL model is reduced to solving the Ermakov equation \cite{Lewis1969} with frequency $\Omega(p,t)$, 
\begin{align}
\ddot{\gamma}_{p}+\Omega^{2}(p,t)\gamma_{p}&=\frac{\omega^{2}_{0p}}{\gamma_{p}^{3}}, \label{eq:Ermakov}\\
\Omega^{2}(p,t)&=\omega_{0p}[\omega_{0p}+2g(p,t)]\label{eq:frequencyErmakov}=\epsilon^{2}(p,t),
\end{align}
The frequency $\Omega(p,t)$ is thus set by the time-dependent spectrum of the TLL. As a consequence, using a state prepared in an interacting excited state $|\{n_{p}\},0\rangle=\bigotimes_{p}|n_{p},0\rangle$ with $\gamma_{p}(0)\neq1$, one finds the evolution of the eigenstates of the TLL Hamiltonian (\ref{eq:eigenstate_evolution}) in the form
\begin{eqnarray}
|\{n_{p}\},t\rangle &=&e^{-\frac{i}{\hbar}\int_{0}^{t}\sum_{p\neq 0}\frac{\epsilon(p,0)}{\sigma^{2}_{s}}\left(n_{p}(0)+\frac{1}{2}\right) ds}Q(t)|\{n_{p}\},0\rangle.\nonumber\\
\label{eq:dynamical_state}
\end{eqnarray}
Here,  $n_{p}(0)$ denotes the occupation number, and $\epsilon(p,0)=\hbar\sqrt{[\omega(p,0)]^{2}-[g(p,0)]^{2}}$ is the $p$-mode energy eigenvalue at the initial time. Note that when the initial state is the ground state of the noninteracting Hamiltonian $|\Omega\rangle$, then  $|\{n_{p}\},0\rangle=|\Omega\rangle$.

\subsection{General solution for the Ermakov equation}
\begin{figure}
  \centering
  \includegraphics[width=1\columnwidth]{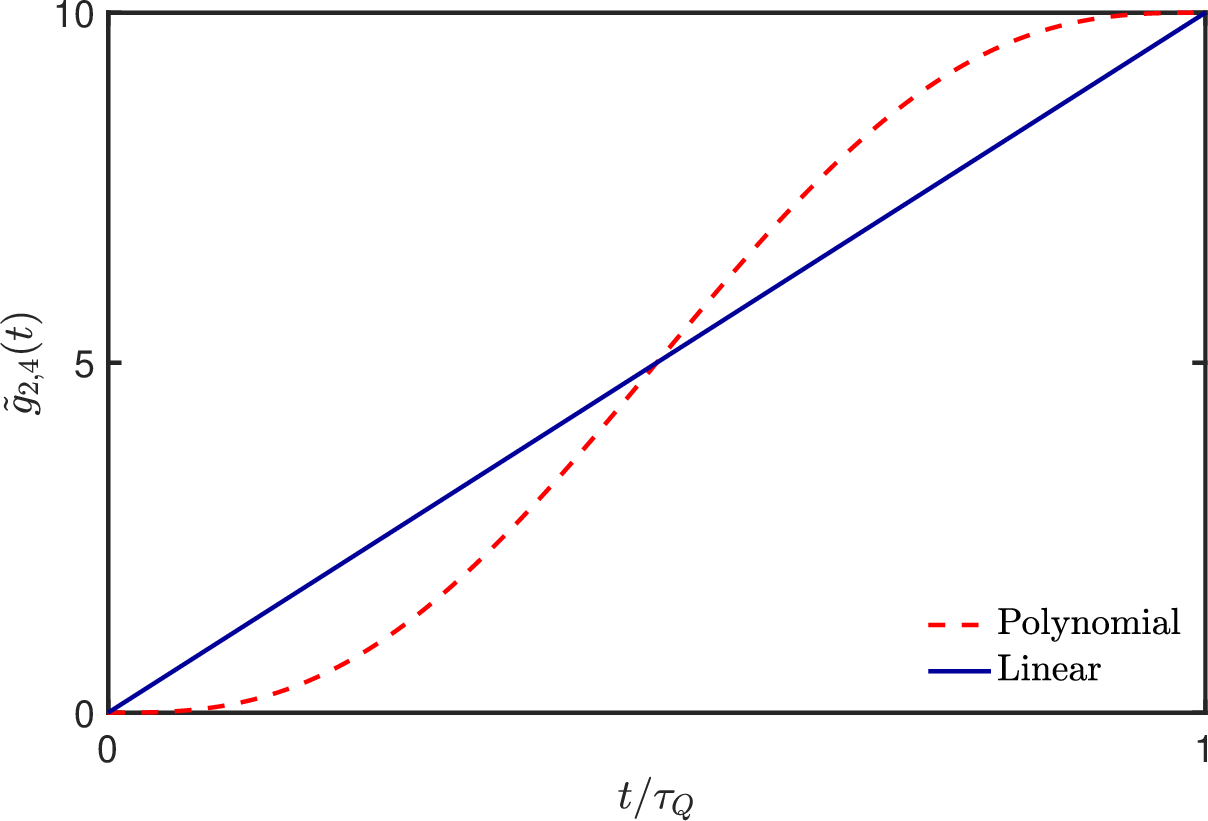}
  \caption{Modulation of the interaction strength from the non-interacting to the interacting regimes in the TLL, during a linear ramp and a smooth polynomial protocol of duration $\tau_{Q}$. We chose the initial state as a non-interacting TLL with $\tilde{g}_{2,4}(0)=0$ and the final interaction strength $\tilde{g}_{2,4}(t)(\tau_{Q})=\alpha=10v_{F}$.}
  \label{fig:modulation_interaction}
\end{figure}

We have seen that it is possible to specify the TLL evolution by solving the Ermakov equation (\ref{eq:Ermakov}), with a frequency depending on the momentum $p$. This equation can be solved numerically for each mode in a forward approach. To solve it analytically, one can rely on the Pinney method \cite{Pinney50}, which consists in writing the solution in the form
\begin{eqnarray}
\gamma_{p}(t)=\sqrt{u^{2}(t)+\frac{\omega^{2}_{0p}}{(W[u,v])^{2}}v^{2}(t)}, \label{eq:pinney_solution}
\end{eqnarray}
where $u$ and $v$ are independent solutions to the homogeneous differential equation $\ddot{\gamma}_{p}+\Omega^{2}(p,t)\gamma_{p}=0$ that verify the initial conditions $u(0)=\gamma_{p}(0),\dot{u}(0)=\dot{\gamma}_{p}(0),v(0)=0,\dot{v}(0)\neq0$, with the corresponding Wronskian being defined as $W[u,v]=u\dot{v}-v\dot{u}$. It is convenient to make further assumptions to simplify the solution. As detailed in App.~\ref{AppPinney}, one can choose $\dot{v}(0)=1/\gamma_{p}(0)$ so that $W[u,v]=1$ and $\dot{\gamma}(0)=0$, leading to the general expression for $u$ and $v$,
\begin{eqnarray}
u(t)&=&\frac{\gamma_{0}}{W[s,r]}[\dot{r}(0)s(t)-\dot{s}(0)r(t)], \label{eq:u_general}\\
v(t)&=&\frac{1}{W[s,r]\gamma_{0}}[s(0)r(t)-r(0)s(t)],\label{eq:v_general}
\end{eqnarray}
where $r$ and $s$ are linearly independent solutions of the homogeneous differential equation and $\gamma_{p}(0)=\gamma_{0}$ for short. The Wronskian of the two independent solutions remains a constant since $\dot{W}[r,s]=0$, which greatly simplifies the form taken by $u$ and $v$. Equation \eqref{eq:pinney_solution} is due to  Pinney \cite{Pinney50}.  The derived equations  (\ref{eq:u_general}) and (\ref{eq:v_general}) are useful in determining an analytical solution for any frequency modulation $\Omega^{2}(p,t)$ for which a solution to the differential equation $\ddot{\gamma}_{p}+\Omega^{2}(p,t)\gamma_{p}=0$ is known. For instance, the homogeneous second-order differential equation can be solved for a constant frequency \cite{Stefanatos10}. For a linear modulation, the equation admits solutions in terms of the Airy functions, and for a quadratic modulation, the solutions can be found in terms of the parabolic cylinder functions \cite{abramowitz}. Other solutions are known \cite{KimKim16,Jaramillo16,Beau16}. For a general time dependence of the frequency, one may find approximated solutions using a perturbative method, such as the WKB approximation.
In the following sections, we discuss the case of linear and smooth modulations of the interaction strength, depicted in Fig.~\ref{fig:modulation_interaction}. 
\subsection{Dynamics following a linear ramp of interactions: Analytical solution via the Ermakov equation \label{section_linear}}
For a contact interaction potential satisfying $g_{2,4}(x-x',t)=\tilde{g}_{2,4}(t)\delta(x-x')$, one finds in the momentum representation that $g_{2,4}(p,t)=\tilde{g}_{2,4}(t)$, i.e., the coupling strengths become momentum independent. We focus on this case and consider the linear ramp of the interaction strength
\begin{eqnarray}
\frac{\tilde{g}_{2,4}(t)}{2\pi}&=&\alpha \frac{t}{\tau_{Q}},
\label{glinramp}
\end{eqnarray}
from the non-interacting regime at $t=0$ to $t=\tau_{Q}$, where $\tau_Q$ sets the ramp duration to reach the interaction strength $\alpha$. This corresponds to the frequency modulation in the Ermakov equation $\Omega^{2}(p,t)=v_{F}[v_{F}+2\alpha t/\tau_{Q}]|p|^{2}$. For this frequency modulation, the homogeneous second-order differential equation takes the form of an Airy differential equation \cite{abramowitz}
\begin{figure}
  \centering
  \includegraphics[width=1\columnwidth]{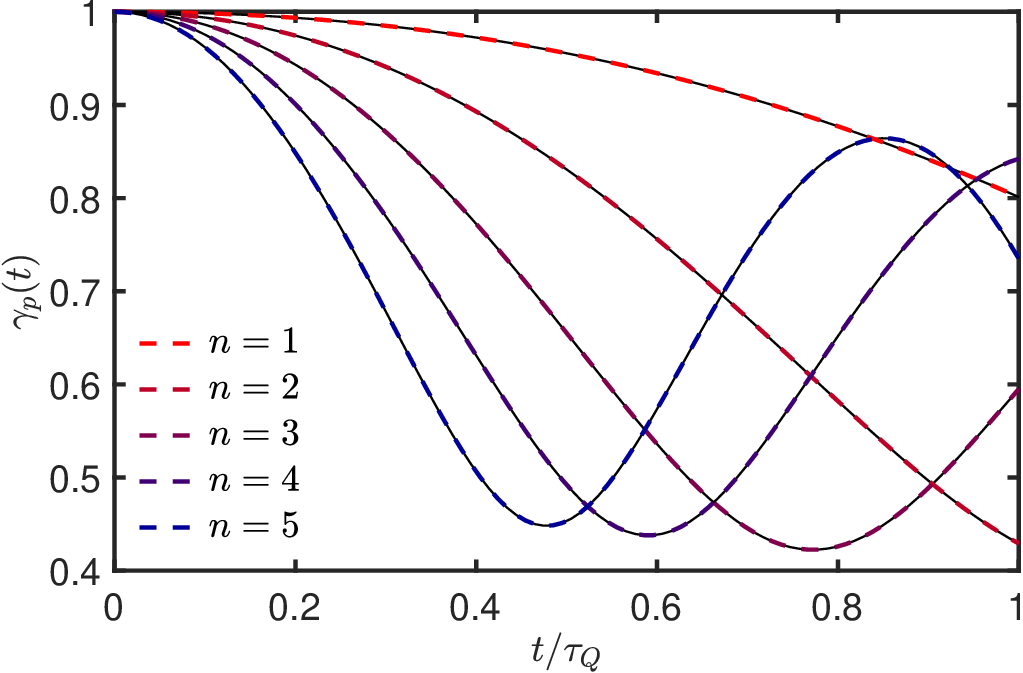}
  \caption{Comparison of the analytical solution to the Ermakov equation with the numerical solution for a linear ramp.  Different modes $p=\frac{2\pi}{L}n$ are considered with $L=100\nu$, the cutoff parameter  $\nu=\hbar/(m v_{F})$ and the frequency modulation $\Omega^{2}(p,t)=v_{F}[v_{F}+2\alpha \frac{t}{\tau_{Q}}]|p|^{2}$. Different modes display an oscillatory behavior that depends distinctively on $n$. The analytical expression (dashed line) exactly matches the numerical solution, shown as a solid black line. The  parameters are $v_{F}=1$ and $\alpha=0.5v_{F}$, with the quench duration  $\tau_{Q}=10$.
  \label{fig:fig_rho}}
\end{figure}
\begin{eqnarray}
\ddot{\gamma}_{p}+\left(at +b\right)\gamma_{p}&=&0 \label{eq:homogeneous},\\
a&=&2v_{F}\alpha p^{2}/\tau_{Q},\label{eqfora}\\
b&=&v^{2}_{F}p^{2},
\end{eqnarray}
whose solutions can be expressed in terms of the Airy functions \cite{abramowitz}, as described in App.~\ref{AppAiry}. The solution to the Ermakov equation (\ref{eq:Ermakov}) with the initial conditions $u(0)=\gamma_{0},\dot{u}(0)=\dot{\gamma}_{p}(0),v(0)=0,\dot{v}(0)=1/\gamma_{0}$ can be expressed as
\begin{figure*}
  \centering
  \includegraphics[width=1.6\columnwidth]{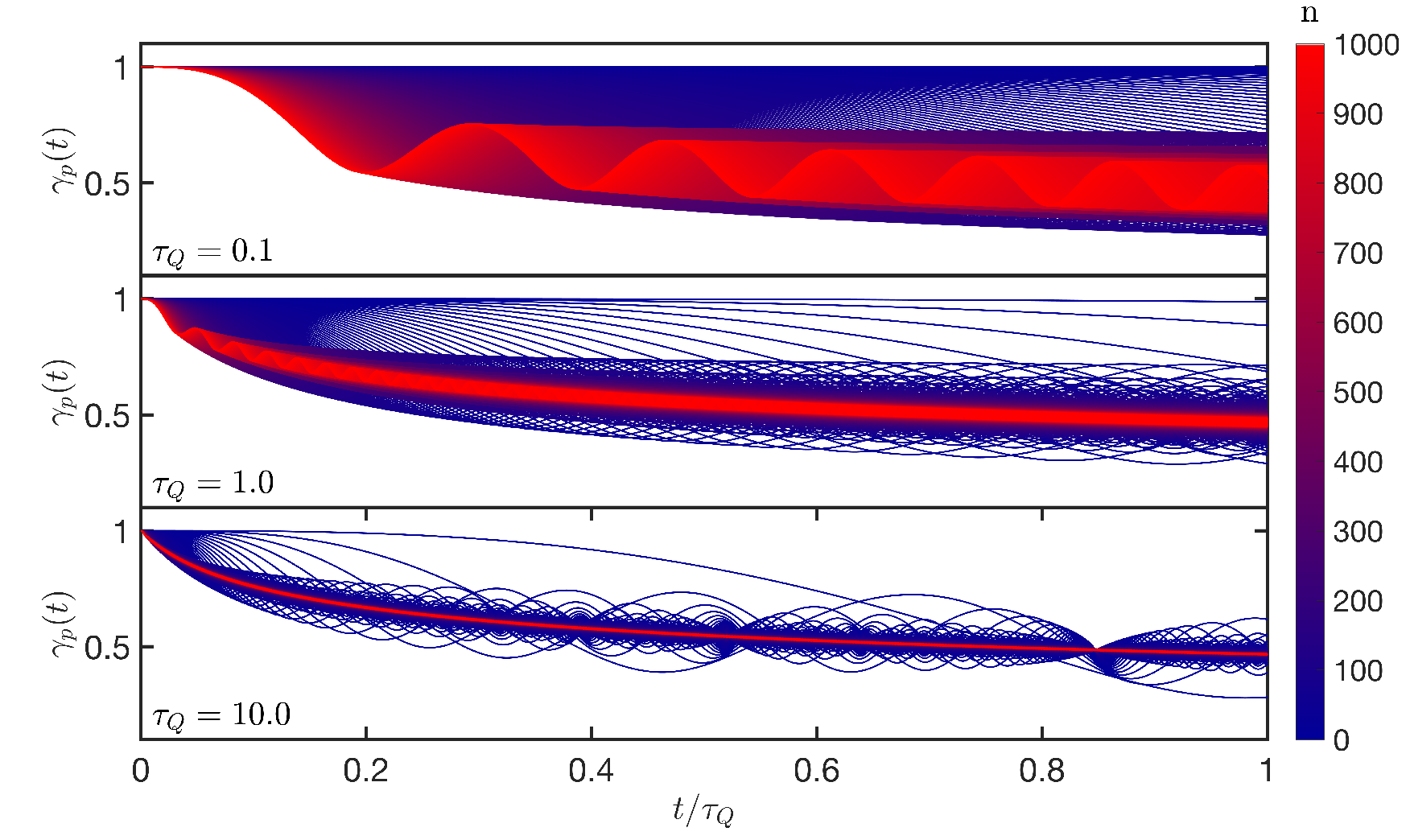}
  \caption{Evolution of the solution to the Ermakov equation for different modes with $p=\frac{2\pi}{L}n$ with $n=1,\cdots,1000$, $L=100\nu$ and cutoff parameter $\nu=\hbar/(m v_{F})$.  The panels correspond to different quench times $\tau_{Q}=0.1,1,10$ when the dynamics is generated by a linear ramp of the interaction strength, leading to a frequency modulation $\Omega^{2}(p,t)=v_{F}[v_{F}+2\alpha \frac{t}{\tau_{Q}}]|p|^{2}$. The solutions develop a large amplitude oscillatory behavior for short quenches for all modes. At low momenta, modes exhibit large amplitude oscillations when the quench duration increases, while the high momenta display a more restricted amplitude for a longer quench. This is consistent with the intuitive picture that only the low-momentum modes are excited for a slow quench. The following parameters are chosen $v_{F}=1$ and $\alpha=10 v_{F}$. One important feature to notice is that the value $\gamma_{p}(\tau_{Q})$ at finite time strongly depends on the quench duration. This explicitly demonstrates that the final state after the quench is not the adiabatic state.
  \label{fig:fig_rho_different_TQ}}
\end{figure*}
{\small
\begin{align}
\begin{split}
z(t)&=-\frac{at+b}{(-a)^{2/3}},\\
u(t)&=\pi\gamma_{0}\{\partial_{z}{\rm Bi}[z(0)]{\rm Ai}[z(t)]-\partial_{z}{\rm Ai}[z(0)]{\rm Bi}[z(t)]\},\\
v(t)&=\frac{\pi}{\gamma_{0}(-a)^{1/3}}\{-{\rm Bi}[z(0)]{\rm Ai}[z(t)]+ {\rm Ai}[z(0)] {\rm Bi}[z(t)]\},\\
\gamma_{p}(t)&=[u^{2}(t)+\omega^{2}_{0p}v^{2}(t)]^{1/2}.\label{eq_Airy_Ermakov}
\end{split}
\end{align}}\normalsize
Figure~\ref{fig:fig_rho} provides a numerical check of the analytical solution, validating their agreement. Figure~\ref{fig:fig_rho_different_TQ} depicts the solution to the Ermakov equation $\gamma_{p}(t)$ for different modes $p$ and compares the time evolution for different quench times with fixed final interaction strength, corresponding to the modulation of $\tilde{g}_{2,4}(t)$ in Eq. (\ref{glinramp}). The behavior of the solution to the Ermakov equation strongly depends on the ramp duration.  For fast ramps, all modes exhibit highly oscillatory behavior. Intermediate ramp timescales reduce the amplitude of the oscillations for highly excited modes. Finally, as the adiabatic limit is approached, the highly excited modes tend to oscillate with restricted amplitude around a defined averaged value. 
 
The adiabatic condition identified by Lewis and Riesenfeld \cite{Lewis1969} sets a  lower bound on the adiabatic time $t_{\rm ad,p}$ for each mode $p$, 
\begin{eqnarray}
\frac{\dot{\Omega}(p,t_{\rm ad,p})}{\Omega^{2}(p,t_{\rm ad,p})}=\frac{2v_{F}\alpha |p|^{2}}{ \tau_{Q}[v_{F}(v_{F}+2\alpha t_{\rm ad,p}/\tau_{Q})|p|^{2}]^{3/2}}\ll1.
\end{eqnarray}
In the limit of slow quenches, whenever $\tau_{Q}\approx t_{\rm ad,p}$, one can obtain a lower bound on the adiabatic time that is roughly proportional to the inverse of the energy of the level. With respect to the quench strength, the length $L$, and Fermi velocity $v_{F}$, the bound can be written in the perturbative limit: 
\begin{eqnarray}
\frac{ L\alpha}{ n\pi v^{2}_{F}}\ll t_{\rm ad,p}\label{eq:adiabatic_time}.
\end{eqnarray}
For $n=1$, the adiabatic time scale is roughly proportional to the inverse of the level spacing. 

We characterize the nonadiabatic character of the evolution via the expectation value of the Hamiltonian.
To this end, we compute the mean energy in App.~\ref{eq:mean_energy} and depict in Fig.~\ref{fig:heat_different_perturbative} the non-adiabatic residual energy after a quench for different values of the interaction strength. Specifically, we consider the non-adiabatic residual energy during the quench, given by the difference between the mean energy in a finite time quench and an infinitely slow protocol. 
We observe an increase in the non-adiabatic residual energy with the strength of the quench. 
\subsubsection{Perturbative solution for the non-adiabatic residual energy during a finite-time quench \label{section_perturbative}}
Exact solutions of many-body quantum systems away from equilibrium are a coveated goal. Apart from their intrinsic interest, they are useful for benchmarking approximate methods.
In what follows, we resort to a perturbative analysis and compare it to the exact solution. In doing so, we gain additional insights into the dynamics following a linear ramp of the interactions and showcase the importance of the exact approach with respect to the perturbative one. 

 It is possible to solve the homogeneous equation (\ref{eq:homogeneous}) for $u$ and $v$ using perturbation theory to first order in the parameter $a$ in Eq. (\ref{eqfora}), as detailed in App.~\ref{perturbative_airy}. This parameter is proportional to the dimensionless parameter $\alpha$ setting the rate of the ramp.
\begin{figure}
\centering
\includegraphics[width=\linewidth]{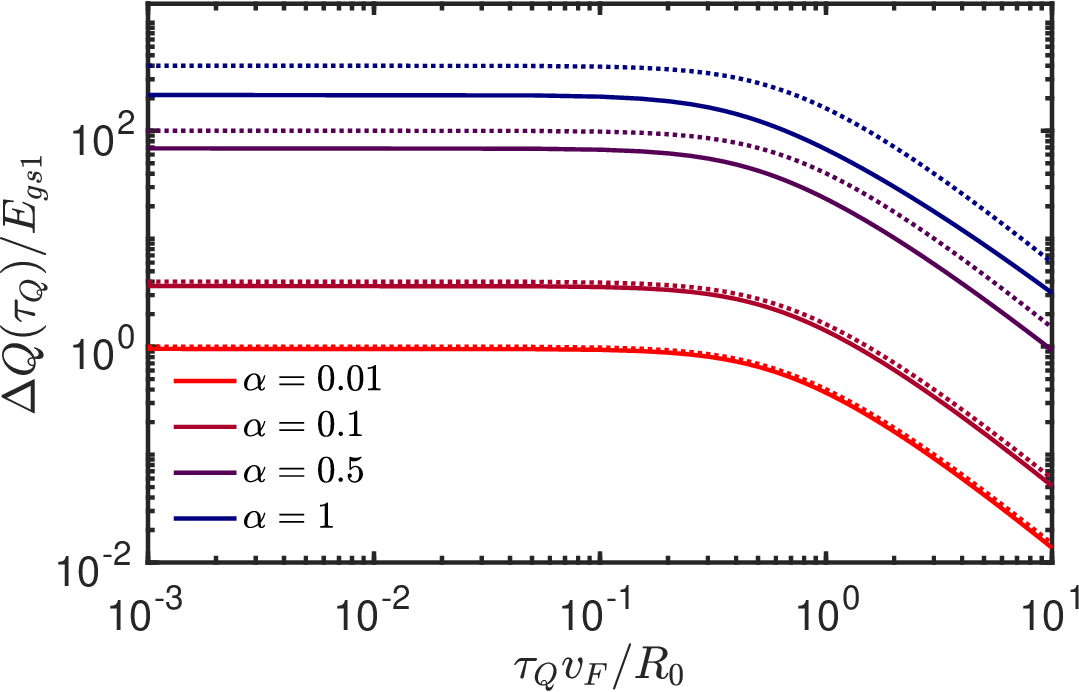}
\caption{Mean non-adiabatic residual energy  $\Delta Q(\tau_{Q})/E_{gs1}$ after a quench of duration $\tau_{Q}$ for different values of the interaction strength after a linear ramp. The energy shift $E_{gs1}=\frac{L}{2\pi}\frac{\hbar \alpha^{2}_{1}}{2v_{F}R^{2}_{0}}$ is chosen as a energy unit. Intuitively, the shorter the quench, the higher the non-adiabatic residual energy is at the end of the protocol. Increasing the interaction strength leads to larger values of the non-adiabatic residual energy. We considered the values $v_{F}=1$, $L=100\nu$ with  $\nu=\hbar/(m v_{F})$, $\gamma_{p}(0)=1$ and $\hbar=1$, $E_{F}\equiv p^{2}_{F}/2m =1/2$, $R_{0}=1$. The exact solution is displayed (solid line) and compared to the perturbative solution for small quenches given by Eq. (\ref{eq:deltaQ_Ermakov}) (dashed line). The perturbative solution is accurate for small quenches values $\alpha \ll v_{F}/2$. For large $\alpha=v_{F}$, there is a mismatch between the exact solution and the perturbative solution associated with the breakdown of the perturbation theory.}
\label{fig:heat_different_perturbative}
\end{figure}
Using the perturbative solution, one can express the non-adiabatic residual energy after the quench as the difference between the finite time mean energy and the adiabatic mean energy
\begin{align}
\Delta Q (\tau_{Q})&=\langle \mathcal{H}(\tau_{Q})\rangle-\langle \mathcal{H}(\tau_{Q}\to \infty)\rangle.
\end{align}
The exact adiabatic energy is computed in App.~\ref{perturbative_airy} for a pure state and can  be Taylor expanded for small $\alpha \ll v_{F}/2$, leading to
\begin{align}
\langle \mathcal{H}(\tau_{Q}\to \infty)\rangle\frac{2\pi}{L}&=\frac{\hbar}{R^{2}_{0}}\left(\sqrt{v_{F}(v_{F}+2\alpha)}\right)\nonumber\\
&\approx \frac{\hbar}{R^{2}_{0}}\left(v_{F}+\alpha-\frac{1}{2}\frac{\alpha^{2}}{v_{F}}\right).
\end{align}
Similarly, for an infinitely short quench, the mean energy is given by
\begin{align}
\langle \mathcal{H}(\tau_{Q}\to 0)\frac{2\pi}{L}=\frac{\hbar}{R^{2}_{0}}(v_{F}+\alpha).
\end{align}
Hence, the non-adiabatic residual energy by an instantaneous quench solely involves a shift in the ground state energy, which scales as $E_{gs}=\Delta Q(0)=\langle \mathcal{H}(\tau_{Q}\to 0)\rangle-\langle \mathcal{H}(\tau_{Q}\to \infty)\rangle\approx\frac{L}{2\pi} \hbar\frac{\alpha^{2}}{2v_{F}R^{2}_{0}}$, in the perturbative limit. For a finite time $\tau_{Q}$, in the thermodynamic limit and for a pure state, the following scaling holds
\begin{align}
\Delta Q(\tau_{Q})&=E_{gs}\left(\frac{\tau_{0}}{\tau_{Q}}\right)^{2}\ln\left[1+\left(\frac{\tau_{0}}{\tau_{Q}}\right)^{2}\right], \label{eq:deltaQ_Ermakov}
\end{align}
with $\tau_{0}=R_{0}/(2v_{F})$. For a short quench time $\tau_{Q}\ll \tau_{0}$, it reduces to
\begin{align}
\Delta Q(\tau_{Q})\approx E_{gs}\left[1-\frac{1}{2}\left(\frac{\tau_{Q}}{\tau_{0}}\right)^{2}\right].
\end{align}
By contrast, in the long quench limit $\tau_{Q}\gg \tau_{0}$,
\begin{align}
\Delta Q(\tau_{Q})\approx 2 E_{gs}\left(\frac{\tau_{0}}{\tau_{Q}}\right)^{2}\ln\left(\frac{\tau_{Q}}{\tau_{0}}\right),
\end{align}
recovering exactly the perturbative result previously obtained in the Heisenberg picture \cite{Dora2011}. In Fig.~\ref{fig:heat_different_perturbative}, the analytical result for the non-adiabatic residual energy is compared to the perturbative solution for different quench strengths. The perturbative solution fits well the curve of the non-adiabatic residual energy for small quenches $\alpha \ll v_{F}/2$. As a consequence, the exact solution we report is particularly interesting in the large quench limit $\alpha \gg v_{F}/2$ and yields lower values of the non-adiabatic residual energy than those predicted by the perturbative solution, that is no longer reliable.
\subsection{Numerical approach to a smooth quench protocol \label{section_polynomial}}
\begin{figure}
  \centering
  \includegraphics[width=1\columnwidth]{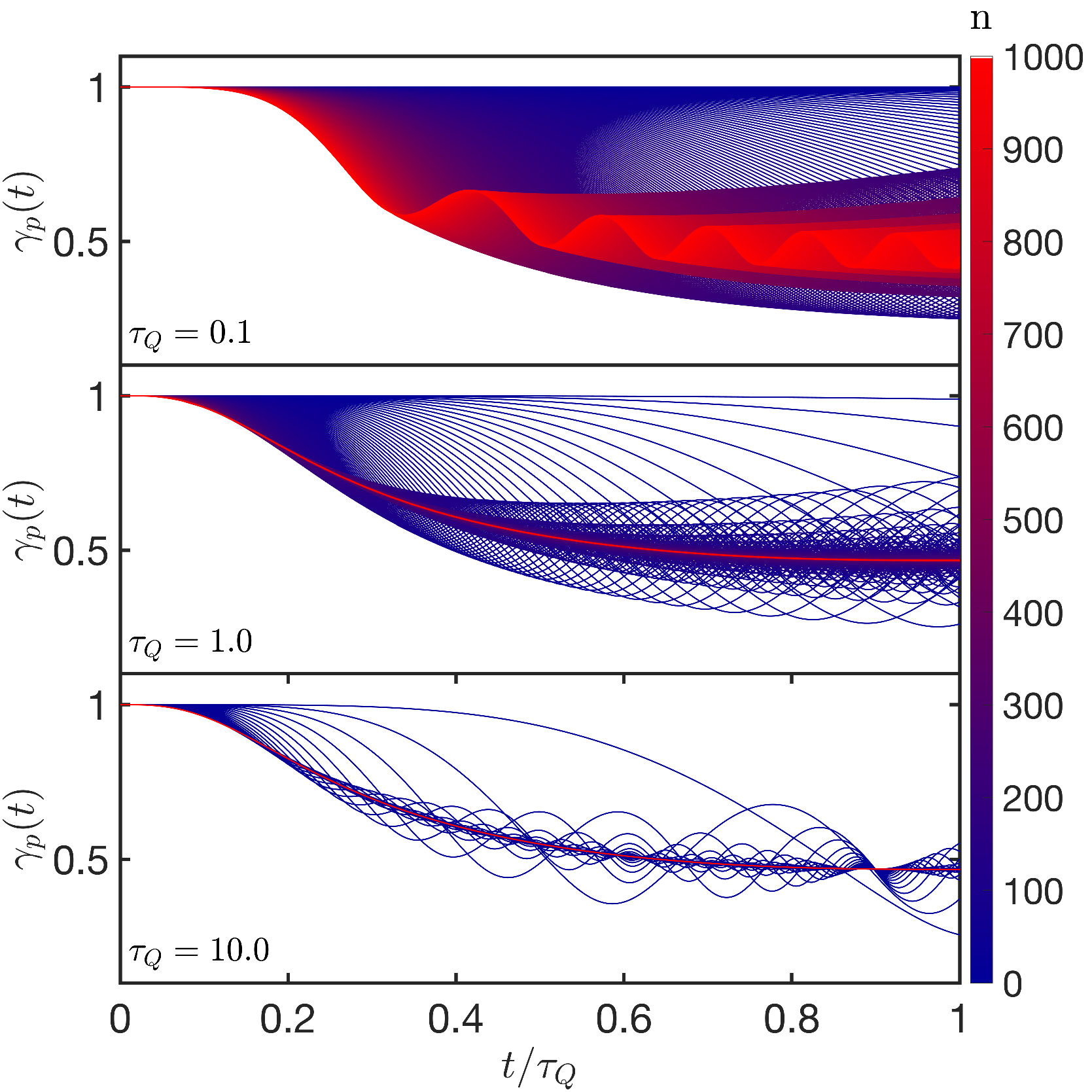}
  \caption{Evolution of the solution to the Ermakov equation for different modes $p=\frac{2\pi}{L}n$ with $n=1,\cdots,1000$ and $L=100\nu$ with  $\nu=\hbar/(m v_{F})$ and different quench times $\tau_{Q}=0.1,1,10$ for a protocol polynomial in time, corresponding to the frequency modulation $\Omega^{2}(p,t)=v_{F}[v_{F}+2\alpha\mathscr{P}(t/\tau_{Q})]|p|^{2}$. The solutions develop a large-amplitude oscillatory behavior for short quenches for all modes, which is particularly prominent at low momenta. The amplitude of such oscillations decreases with the quench time. The following parameters are chosen $v_{F}=1$ and $\alpha=10 v_{F}$.
  \label{fig:fig_rho_different_TQ_fifth_order}}
\end{figure}

\begin{figure}
  \centering
  \includegraphics[width=1\columnwidth]{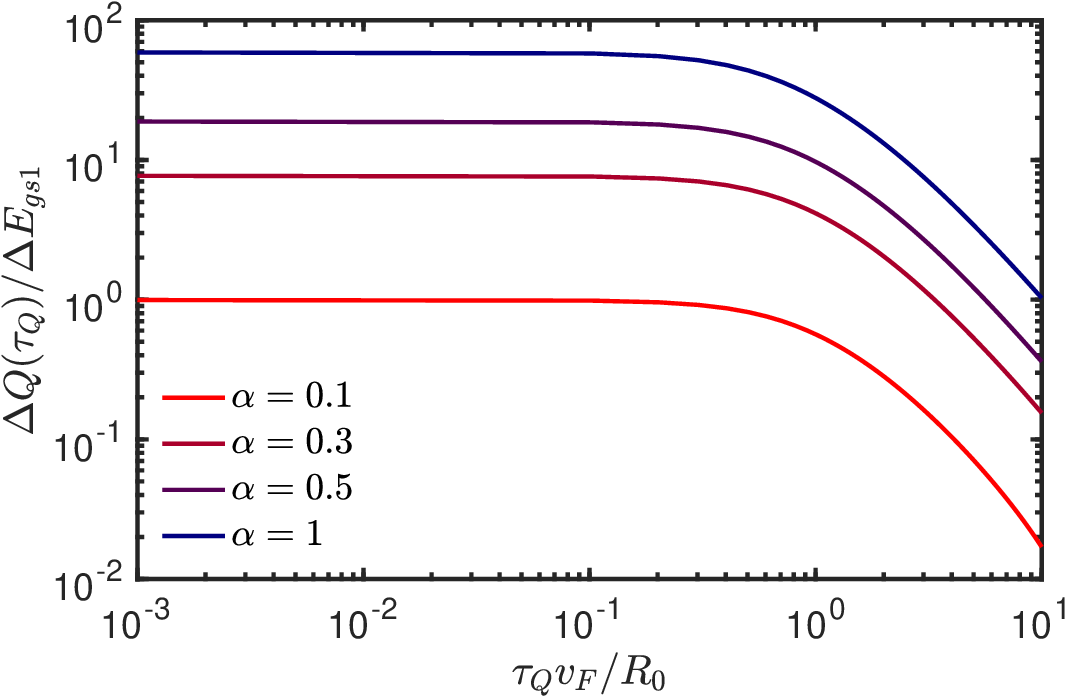}
  \caption{Mean non-adiabatic residual energy after a smooth interaction quench described by a fifth-order polynomial $\Delta Q(\tau_{Q})/E_{gs1}$ where $\Delta E_{gs1}=\frac{L}{2\pi}\frac{\hbar(v_{F}+\alpha_{1})}{R^{2}_{0}}-\frac{L}{2\pi}\frac{\hbar \sqrt{v_{F}(v_{F}+2\alpha_{1})}}{R^{2}_{0}}$. Shortening the quench increases the non-adiabatic residual energy. Similarly, the non-adiabatic residual energy increases with the interaction strength. We considered the values $v_{F}=1$, $L=100\nu$ with $\nu=\hbar/(m v_{F})$ the cutoff parameter and, $\gamma_{p}(0)=1$ and $\hbar=1$, $E_{F}=1/2$, $t_{\rm ad}=100$ and, $R_{0}=1$. We note that the behavior of the non-adiabatic residual energy for the smooth polynomial is similar to the linear quench driving. Similarly, we observe a plateau at a short time, followed by a smooth decay of the non-adiabatic residual energy in the adiabatic limit.}
  \label{fig:fig_res_mean_energy_fifth_order_TQ}
\end{figure}

When analytical solutions are unavailable, an approach relying on the Ermakov equation is still fruitful. By way of example, we numerically investigate the dynamics generated by a smooth driving protocol of the coupling strength and frequency. Specifically, for the interaction strength $\tilde{g}_{2}(t)=\tilde{g}_{4}(t)=\tilde{g}_{2,4}(t)$, we consider a modulation in time following a fifth-order polynomial of the form 
\begin{eqnarray}
\frac{\tilde{g}_{2,4}(t)}{2\pi}&=&\alpha\mathscr{P}(t/\tau_{Q}),\\
\mathscr{P}(s)&=&10 s^{3}-15 s^{4}+6 s^{5}.
\end{eqnarray}
This provides an associated frequency for the Ermakov equation
\small
\begin{eqnarray}
\Omega^{2}(p,t)=v_{F}[v_{F}+2\alpha\mathscr{P}(t/\tau_{Q})]|p|^{2}. \label{eq:omega_fifth}
\end{eqnarray}\normalsize
Figure~\ref{fig:fig_rho_different_TQ_fifth_order} represents the numerical solution to the Ermakov equation for the fifth-order polynomial quench. As in the linear quench, the amplitude of the oscillations varies with the quench duration and the higher modes converge to the adiabatic solution for a large quench duration. Figure~\ref{fig:fig_res_mean_energy_fifth_order_TQ} pictures the non-adiabatic residual energy following the frequency modulation in Eq. (\ref{eq:omega_fifth}) for different quench strength $\alpha$ starting from a noninteracting state.
%
\subsection{Exact Solution for an Inverse-Polynomial Quench \label{section_polynomial_quench}}
Before closing, we  discuss a class of solutions to the Ermakov equation for which $\ddot{\gamma}_{p}=0$ at all times. We show that it is possible to determine a driving protocol that verifies this constraint. Let us consider the frequency in  the Ermakov equation corresponding to a delta-interacting quench $g_{2,4}(x-x',t)=\tilde{g}_{2,4}(t)\delta(x-x')$, i.e., 
\begin{eqnarray}
\Omega^{2}(p,t)=v_{F}\left[v_{F}+2\frac{\tilde{g}_{2,4}(t)}{2\pi}\right]|p|^{2}.
\label{LorentzOm}
\end{eqnarray}
Setting $\ddot{\gamma}_{p}=0$ in the Ermakov equation, one can recast the frequency in the form $\Omega^{2}(p,t)=\Omega^{2}(p,0)/\gamma^{4}_{p}$ with $\gamma_{p}=(At^{2}+2Bt+C)^{1/2}$,  provided that the condition $\Omega^{2}(p,0)+AC-B^{2}=\omega^{2}_{0p}$ is satisfied \cite{Berry_1984,Dodonov92,Rezek2009}. Using the boundary conditions $\tilde{g}_{2,4}(0)=0$ and $\tilde{g}_{2,4}(\tau_{Q})$, we can determine the values for $A,B$ and $C$ so that $AC=B^{2}$ and $C=1$, which  yields $B=\frac{1}{\tau_{Q}}\left[-1\pm\left(\frac{\tilde{g}_{2,4}(\tau_{Q})}{\pi v_{F}}+1\right)^{-1/4}\right]$. We note that in the adiabatic limit $\tau_{Q}\to +\infty$, $ B \to 0$. Finally, the solution to the Ermakov equation simplifies to $\gamma_{p}=Bt+1$ and the driving protocol in (\ref{LorentzOm}) takes the form
\normalsize
\begin{eqnarray}
\label{eq:lambda}
\frac{\tilde{g}_{2,4}(t)}{2\pi}=\frac{v_{F}}{2}\left[\frac{1}{(Bt+1)^{4}}-1\right].
\end{eqnarray}\normalsize
Note that this protocol follows from reverse-engineering the dynamics, as it involves prescribing a modulation of $\gamma_p(t)$ and subsequently determining the associated required modulation of the interaction strength. As a consequence, for any duration of the quench $\tau_{Q}$ the final value of the solution of the Ermakov equation is fixed to $\gamma_{p}(\tau_{Q})= \left(\frac{\tilde{g}_{2,4}(\tau_{Q})}{\pi v_{F}}+1\right)^{-1/4}$. We compute the mean energy in App.~\ref{eq:mean_energy} and determine the non-adiabatic residual energy for the protocol given by Eq. (\ref{eq:lambda}) in the thermodynamic limit for a pure state, to be given by
\begin{align}
\Delta Q(\tau_{Q})&=\frac{\hbar B^{2}}{2}\frac{L}{2\pi}\frac{1}{v_{F}}\Gamma(0,\frac{2\pi}{L}R_{0}),
\end{align}
where the incomplete gamma function is defined as $\Gamma(s,x)=\int_{x}^{\infty}t^{s-1}e^{-t}dt$. In particular, in the small quench perturbative limit $\frac{\tilde{g}_{2,4}(t)}{2\pi}\ll v_{F}/2$
\begin{align}
\Delta Q(\tau_{Q})&\approx  E_{gs} \left(\frac{\tau_{0}}{\tau_{Q}}\right)^{2} \Gamma(0,\frac{2\pi}{L}R_{0}),\label{eq:perturbative_reverse_engineering}
\end{align}
with $E_{gs}=\frac{\hbar L}{2\pi}\frac{(\tilde{g}_{2,4}(\tau_{Q})/2\pi)^{2}}{2v_{F} R^{2}_{0}}$. This result differs from the previous protocols and shows that the dependence on the quench time can be tailored by modifying the protocol.  
Specifically, the logarithmic dependence of the quench time is removed for the inverse-polynomial quench protocol.

\section{Summary and Outlook \label{section_summary}}
Exact solutions in nonequilibrium physics are highly valuable for the insights they unveil and as a benchmark for numerical methods, approximate schemes, and experiments, e.g., in quantum simulation and condensed matter physics. They are primarily available in the adiabatic limit or sudden quenches.

We have reported the exact description of the quantum dynamics of a TLL following an arbitrary interaction modulation in time. 
To this end, we have used the Schr\"odinger picture and exploited the $SU(1,1)$ dynamical symmetry group.
We have shown how nonadiabatic protocols tailoring a semiclassical sine-Gordon potential can be engineered to perform an STA, leading to the preparation of a final state without residual excitations. 

We have further shown that this framework makes possible the analytical description of the TLL  driven in finite time, as we have shown by characterizing the nonequilibrium dynamics resulting from interaction quenches that involve a linear or polynomial schedule in time.

Our results should find broad applications in the nonequilibrium physics of low-dimensional quantum systems and effective field theories, e.g., in the study of thermalization, ergodicity, adiabaticity, and their shortcuts. Our work paves the way to characterize the TLL dynamics induced by finite-time driving protocols, e.g., via work statistics and Loschmidt echoes. The development of these STA protocols driving nonadiabatically between gapless phases constitutes an important advance in the control of many-body systems and nonequilibrium quantum field theory \cite{Calzetta08}. These findings are particularly striking in view of the arguments against building on the breakdown of adiabatic continuation in low-dimensional systems \cite{Polkovnikov08}.
It is worth considering whether similar STA protocols can be realized in scenarios of particle production \cite{Birrell82,Dabrowski15,Dabrowski16,Deffner16}.

The exact description of the TLL dynamics can be applied to finite-time thermodynamics, where the realization of quantum heat engines with a TLL as a working substance has been proposed \cite{ChenWatanabe2019,Gluza21}. 
In this context, optimal control can enhance the output power and efficiency of quantum devices \cite{Salamon09,delcampo14,Beau16,Deng18Sci,delCampo2018}. For instance, in a quantum Otto cycle such as the one described in \cite{ChenWatanabe2019}, the adiabatic strokes of the heat engine could be enhanced by the shortcut in tuning both the short-range interaction strength and an external lattice potential as suggested in section \ref{lattice_implementation}. An alternative implementation resides in a long-range $1D$ Coulomb interaction as described in section \ref{long_range}. Combining finite-time driving with spatially inhomogeneous protocols \cite{Gluza22,moosavi2023exact} has proved helpful in reducing excitations in quantum critical systems \cite{Collura10,Dziarmaga10,GomezRuiz19} and may facilitate the quantum control of the TLL dynamics. Developing optimal control protocols for the TLL may be assisted by using the generalization of delta kick cooling for scale-invariant quantum fluids \cite{Dann19,dupays2021,Dupays23}.  A challenging prospect involves the generalization of our results to nonunitary processes such as cooling and heating \cite{Dupays2020,Alipour20}, paving the way for the engineering of fully superadiabatic quantum thermodynamic cycles \cite{delCampo2018,pedram2023quantum} in the many-particle case.  It would also be interesting to compare our analytical results to studies of quenches in the XXZ model \cite{Schoenauer2019} and the Lieb-Liniger model \cite{Calabrese_2014}. Our formalism can be further used to compute correlation functions  \cite{Iucci2009}, which can display nonequilibrium universal behavior. Finally, our formalism relies on the Ermakov equation, which has many known solutions and is easy to solve numerically. As a result, our work should facilitate the study of nonequilibrium quantum dynamics induced by finite-time driving protocols.

\acknowledgements
The authors are indebted to Per Moosavi for his thorough comments on the manuscript. The authors would also like to thank Apollonas S. Matsoukas Roubeas, Jingjun Zhu, Federico Balducci, Thomas Schmidt, and Jing Yang for insightful discussions.  This project was supported by the Luxembourg National Research Fund (FNR Grant No. 16434093). It has received funding from the QuantERA II Joint Programme with co-funding from the European Union’s Horizon 2020 research and innovation programme.
This research was further supported by the Ministry of Culture and Innovation and the National Research, Development, and
Innovation Office within the Quantum Information National Laboratory of Hungary (Grant No. 2022-2.1.1-NL-2022-00004)
K134437, K142179, and by a grant of the Ministry of Research, Innovation and Digitization, CNCS/CCCDI-UEFISCDI, under
project number PN-III-P4-ID-PCE-2020-0277. 
\setcounter{secnumdepth}{2} 
\appendix
\onecolumngrid 
\section{Overview of the Tomonaga-Luttinger model \label{app:TLL}}
In this section we provide an overview of the TLL model, fixing the notation. It describes massless fermions on a domain of length $L$ with Fermi velocity $v_F$. The free Hamiltonian $H_0$ of the TLL model is given by $H_0 = v_F \int_0^L \normord{\Psi^\dagger(x) (\sigma_3 p-k_{F}) \Psi(x)} dx$ \cite{Luttinger1963, Haldane1981}, where $\sigma_3$ is the Pauli matrix, $p = -i\hbar \partial_x$, $k_{F}$ the Fermi momentum and $\normord{\mathcal{O}}$ denotes the operator normal ordering relative to the ground state of the free Hamiltonian. The Dirac equation solution is represented by the spinor $\Psi = (\Psi_1, \Psi_{-1})$, which can be expressed as a Fourier series $\Psi_s(x) = \frac{1}{\sqrt{L}}\sum_k a_{sk} e^{ikx}$, where $s = \pm 1$. Periodic boundary conditions impose quantization of the momentum $k = \frac{2\pi n}{L}$, where $n \in \mathbb{Z}$. The fermionic creation and annihilation operators $a_{sk}^\dagger$ and $a_{sk}$ satisfy the conventional anticommutation relations $\{a_{sk}^\dagger, a_{s'k'}\} = \delta_{ss'}\delta_{kk'}$. In terms of these operators, the free Hamiltonian reads $H_0 = \hbar v_F \sum_{r=\pm 1,k} (rk-k_{F}) \normord{a_{rk}^\dagger a_{rk}}$.  The particles with positive velocity, labeled by `$1$', are referred to as right-movers, while those with negative velocity, labeled by `$-1$', are called left-movers. As noticed by Mattis and Lieb \cite{Mattis1965}, the definition of the ground state and the presence of an infinity of particles in the Dirac sea requires care. It is convenient to express right and left movers in terms of particles and holes \cite{Luttinger1963,Mattis1965}, using $a_{1k}=\Theta(k-k_{F})b_{k}+\Theta(k_{F}-k)c^{\dagger}_{k}$ and $a_{-1k}=\Theta(k+k_{F})c^{\dagger}_{k}+\Theta(-k-k_{F})b_{k}$, in terms of the Heaviside step function, satisfying $\Theta(x)=1 $ for $x>0$ and $\Theta(x)=0$ otherwise.
 In these expressions,  $b^{\dagger}_{k}$ and $b_{k}$ are  creation and annihilation operators for particles, while  $c^{\dagger}_{k}$ and $c_{k}$ are creation and annihilation operators for holes. As a consequence, the ground state is the filled Dirac sea with no particles and holes and can be expressed in the particle-number and hole-number basis as $|\Omega\rangle=|0_{b}\rangle\otimes|0_{c}\rangle$. The expectation value of the number operator for right and left movers $\hat{n}_{s}=a^{\dagger}_{s}a_{s}$ (with $s=\pm 1$) in the vacuum  depends on the momentum as \cite{Schindler2022}
\begin{eqnarray}
\label{eq:ground_a1}
\langle \Omega | a^{\dagger}_{1k}a_{1k}|\Omega\rangle= \begin{cases}
 \langle \Omega | b^{\dagger}_{k}b_{k}|\Omega\rangle=0& \text{for} \quad k>k_{F},\\
 \langle \Omega | c_{k}c^{\dagger}_{k}|\Omega\rangle=1 & \text{for} \quad k<k_{F}.
\end{cases}
\end{eqnarray}
Similarly, $\langle \Omega | a^{\dagger}_{-1k}a_{-1k}|\Omega\rangle= \Theta(k+k_{F})$. For $p>0$ and $s=\pm 1$, one can define the densities \cite{Mattis1965}   $\rho_{s}(p)=\sum_{k}\normord{a^{\dagger}_{s,k+p}a_{s,k}}$ and $ \rho_{s}(-p)=\sum_{k}\normord{a^{\dagger}_{s,k}a_{s,k+p}}$. In particular, for $p=0$, $\rho_{s}(0)=\sum_{k}\normord{a^{\dagger}_{sk}a_{sk}}=N_{s}$. It is thus possible to express the Hamiltonian of the TLL model in terms of the densities using normal ordering to compute the commutation relation \cite{Mattis1965} for $p'>0$ $[\rho_{1}(-p),\rho_{1}(p')]=[\rho_{-1}(p),\rho_{-1}(-p')]=\frac{p L}{2\pi}\delta_{p,p'}$. These densities can be reformulated in terms of bosonic operators that verify the usual commutation relation $[b(p),b^{\dagger}(p')]=\delta_{p,p'}$  for $p>0$, with $b(p)=\sqrt{\frac{2 \pi}{L |p|}}\left[\Theta(p)\rho_{1}(-p)+\Theta(-p)\rho_{-1}(-p)\right]$ and $b^{\dagger}(p)=\sqrt{\frac{2 \pi}{L |p|}}\left[\Theta(p)\rho_{1}(p)+\Theta(-p)\rho_{-1}(p)\right]$. Using the previous identities, the free Hamiltonian is written \cite{Mattis1965}
\begin{eqnarray}
H_{0}&=&\sum_{p\neq 0}v_{F}\hbar |p| b^{\dagger}(p)b(p)+\frac{\pi v_{F}}{L}(N^{2}+J^{2}),
\end{eqnarray}
with $N=N_{1}+N_{-1}$ and $J=N_{1}-N_{-1}$ \cite{Mattis1974}. Eigenvalues of the current $J$ operator depend on the boundary conditions chosen \cite{Cazalilla2004}.

Let us now introduce interactions in the model so that the complete Hamiltonian reads $H_{\rm TL}=H_{0}+V$. Neglecting the chemical potential and interactions that do not conserve the number of right and left movers,  the interaction potential takes the form $\hat{V}=\sum_{i,j=\pm1}\hat{V}_{ij}$ with 
{\small
\begin{align}
\hat{V}_{ij}&=\frac{\pi \hbar}{L}\sum_{p>0}\{V_{ij}(p,t)\rho_{i}(-p)\rho_{j}(p)+V_{ij}(-p,t)\rho_{i}(p)\rho_{j}(-p)\}\nonumber\\&+\frac{\pi \hbar}{L}V_{ij}(0,t)\rho_{i}(0)\rho_{j}(0),
\end{align}}
where
\begin{eqnarray*}
\label{eq:interaction_potential}
V_{ij}(p,p',t)&=&\frac{1}{L}\int_{0}^{L} dx \int_{0}^{L} dx'e^{ipx} V_{ij}(x-x',t) e^{ip'x'},\\
V_{ij}(p,p',t)&=&\delta_{p,-p'}V_{ij}(p,t).
\end{eqnarray*}
Even parity of the interaction potential is assumed in the following, i.e.,  $V_{ij}(p,t)=V_{ij}(-p,t)$.  One can express the densities in the bosonic basis. In particular, the interaction potential between the right and left movers  can now be written in terms of the $\mathfrak{su}(1,1)$ generators as 
\begin{align}
H_{2}=\hat{V}_{1,-1}+\hat{V}_{-1,1}=\frac{\hbar}{2}\sum_{p\neq0}|p|\frac{g_{2}(p,t)}{2\pi}[K_{+}(p)+K_{-}(p)]+\frac{\hbar g_{2}(0,t)}{4 L}(N^{2}-J^{2}),
\end{align}
where  $\frac{g_{2}(p,t)}{2\pi}=V_{1,-1}(p,t)$ is the interaction strength. Similarly,  the interaction potential of right and left movers with themselves is described by \begin{align}
H_{4}=\hat{V}_{-1,-1}+\hat{V}_{1,1}=\hbar \sum_{p\neq0}|p|\frac{g_{4}(p,t)}{2\pi}K_{0}(p)+\frac{\hbar g_{4}(0,t)}{4L}(N^{2}+J^{2}),\end{align}
with  coupling strength $\frac{g_{4}(p,t)}{2\pi}=V_{1,1}(p,t)=V_{-1,-1}(p,t)$. The interaction potentials verify the properties $|\frac{g_{2}(p,t)}{2\pi}|<(\frac{g_{4}(p,t)}{2\pi}+v_{F})$ and $\frac{g_{2}(p,t)}{2\pi}/(v_{F}+\frac{g_{4}(p,t)}{2\pi})\to 0$ as $|p|\to\infty$, faster than $|p|^{-1/2}$ \cite{Haldane1981}. Collecting the free and interacting parts, the TLL Hamiltonian reads
\begin{eqnarray}
H_{\rm TL}(t)=H_{0}+H_{2}+H_{4}. \label{eq:LLLLL}
\end{eqnarray}
The field representation of the TLL is obtained by using the decomposition of the field operator decomposes into the right (denoted $`-1'$) and left movers (denoted $`+1'$) \cite{Giamarchibook}
\begin{align}
\Psi(x)\approx e^{ik_{F}x}\Psi_{1}(x)+e^{-ik_{F}x}\Psi_{-1}(x).
\end{align}
Using the commutation relations between the field and the densities, one can express the right and left movers in the form \cite{Mattis1974, Luther1974}
\begin{eqnarray}
\Psi_{s}(x)&=&\eta_{s}\sqrt{\rho_{0}/2}e^{i s \phi_{s}(x)},\label{eq:bosonized_right_left_mover}
\end{eqnarray}
 where $\eta_{s}$ denote the Klein factors defined in \cite{Delft1998}, that ensure the anticommutation of left and right movers, i.e.,  $\{\eta_{s},\eta_{s'}\}=0$ for $s\neq s'$. Klein factors lower the $s$-fermion number by one according to $[N_{s},\eta_{s'}]=-\delta_{ss'}\eta_{s}$ and similarly raise the $s$-fermion number by one, as a result of $[N_{s},\eta^{\dagger}_{s'}]=\delta_{ss'}\eta^{\dagger}_{s}$. They further satisfy  $\{\eta^{\dagger}_{s},\eta_{s'}\}=2\delta_{ss'}$ with $\eta_{s}\eta^{\dagger}_{s}=\eta^{\dagger}_{s}\eta_{s}=1$. The phase is then written as
\begin{eqnarray}
\phi_{s}(x)&=&\Phi_{s}(x)+\Phi^{\dagger}_{s}(x),\\
\Phi_{s}(x)&=& \lim_{\nu \to 0} -i s\sum_{k>0}\sqrt{\frac{2\pi}{kL}}e^{i s k x}e^{-\nu |k|/2}b(sk).
\end{eqnarray}  
One can equate the expression of the field (\ref{eq:bosonized_field}) near the Fermi momentum and identify the expression for the right and the left movers (\ref{eq:bosonized_right_left_mover}) in order to obtain the expressions for the fields $\theta(x)$ and $\phi(x)$ given in (\ref{eq:phi_expression}) and (\ref{eq:theta_expression}).
\section{Implementation of the TLL with $g_{2}(p,t)=g_{4}(p,t)$ \label{app:implementation_TLL}}
The general TLL is valid for $g_{2}(p,t)\neq g_{4}(p,t)$. However, some of our results focus on the specific case $g_{2}(p,t)=g_{4}(p,t)$. There are several physical implementations where this scenario applies. For instance, this is of relevance to a 1D Bose quantum fluid with short-range interparticle interactions described by a delta function, as firstly noticed in Haldane's harmonic fluid approach for bosons \cite{Haldane1981PRL,Gritsev2007}, and experimentally studied in \cite{Tajik2023}. The Hamiltonian is then
\begin{eqnarray}
H_{\rm TL}&=&\int_{0}^{L}dx\left(\frac{1}{2m}\partial_{x}\Psi^{\dagger}(x)\partial_{x}\Psi(x)+\frac{g(t)}{2}\Psi^{\dagger}(x)\Psi(x)\Psi^{\dagger}(x)\Psi(x)-\mu\Psi^{\dagger}(x)\Psi(x)\right)\\
&\approx&\frac{v_{s}(t)}{2 \pi}\int_{0}^{L} dx \left[ K(t)(\partial_{x}\phi(x))^{2}+K^{-1}(t)\left(\pi\rho_{0}-\partial_{x}\theta(x)\right)^{2}\right]-\int_{0}^{L} dx
\mu\left(\rho_{0}-\frac{\partial_{x}\theta(x)}{\pi}\right)\\
&=&\frac{v_{s}(t)}{2 \pi}\int_{0}^{L} dx \left[ K(t)(\partial_{x}\phi(x))^{2}+K^{-1}(t)\left(\partial_{x}\theta(x)\right)^{2}\right]-\frac{\rho^{2}_{0}gL}{2},
\end{eqnarray}
where $m$ is the mass of the atoms, $g(t)\geq 0$ is an effective interaction strength, and $\mu$ the chemical potential that verifies $\rho_{0}=\mu/g$. Following the harmonic fluid approach $\Psi^{\dagger}(x)\approx\sqrt{\rho_{0}-\frac{\partial_{x}\theta(x)}{\pi}}e^{-i\phi(x)} $, one finds
\begin{eqnarray}
v_{s}(t)&=&\sqrt{\frac{g(t)\rho_{0}}{m}},\\
K(t)&=&\pi\sqrt{\frac{\rho_{0}}{mg(t)}}. \label{eq:K_ultracold_atoms}
\end{eqnarray}
This yields
$v_{s}(t)/K(t)=g(t)/\pi$, leading to $g(t)=\pi(v_{F}+\tilde{g}_{2,4}(t)/\pi)$. In particular, note that the noninteracting case corresponds to $v_{F}=-\tilde{g}_{2,4}(t)/\pi$, which leads to an infinite Luttinger parameter $K(t)\to +\infty$. 
For the Lieb-Liniger model \cite{LiebLiniger1963}, the dimensionless coupling is defined as $\upsilon=mg/\rho_{0}$. Analytical expressions for $v_{s}(\upsilon)$ and $K(\upsilon)$ are not known. However, approximated expressions can be found in the limit of large and small $\upsilon$ \cite{Buchner2003} when 
\begin{align}
v_{s}=v_{F}K^{-1}, &\quad K\approx \begin{cases} 1+\frac{4}{\upsilon} &\text{for} \quad \upsilon \gg 1, \\
\frac{\pi}{\sqrt{\upsilon}}\left(1-\frac{\sqrt{\upsilon}} {2\pi}\right)^{-1/2} &\text{for}\quad \upsilon \ll 1.
\end{cases}
\end{align}
Hence, we note that in the limit $\upsilon \ll 1$, one recovers the expression (\ref{eq:K_ultracold_atoms}). 

Another example of relevance is the XXZ spin chain. Let us consider the spin $1/2$ chain, in which each site is endowed with a spin degree of freedom $S_{i}^\alpha=\sigma_{i}^\alpha/2$, where $\sigma_{i}^\alpha$ ($\alpha=x,y,z$) denotes the Pauli matrix $\sigma^\alpha$ acting on the $i$-th site. The spin components $S^{x}_{j},S^{y}_{j},S^{z}_{j}$ act on lattice sites $j=1,\cdots,N$ and obey the commutation relations $[S^{\alpha}_{j},S^{\beta}_{j'}]=i\delta_{jj'}\epsilon_{\alpha\beta \gamma}S^{\gamma}$ where $\epsilon_{\alpha \beta \gamma}$ is the totally antisymmetric tensor (satisfying $\epsilon_{xyz}=1$ and zero if two indices are equal). The XXZ Hamiltonian reads 
\begin{eqnarray}
H=\sum_{j=1}^{N}J_{xy}(S^{x}_{j+1}S^{x}_{j}+S^{y}_{j+1}S^{y}_{j})+\sum_{j=1}^{N}J_{z}S^{z}_{j+1}S^{z}_{j},
\end{eqnarray}
and $L=Na$ with $a$ the lattice spacing. The TLL Hamiltonian associated with the XXZ model can be expressed as \cite{Giamarchibook} 
\begin{eqnarray}
H=H_{\rm TL}-2g_{3}\rho^{2}_{0}\int dx \cos[4\theta(x)], \label{eq:bosonized_spin_chain}
\end{eqnarray}
where $g_{3}=a J_{z}$ and $H_{\rm TL}$ is the TLL of Eq. (\ref{eq:field_TLL}) with parameters $v_{s}(t)K(t)=v_{F}=J_{xy}a\sin(k_{F}a)$ and $v_{s}(t)K^{-1}(t)=v_{F}(1+\frac{2J_{z}a}{\pi v_{F}}[1-\cos(2k_{F}a)])$. Hence, we have $\frac{\tilde{g}_{2,4}(t)}{2\pi}=\frac{J_{z}a}{\pi}[1-\cos(2 k_{F}a)]$ that reduces to $\frac{\tilde{g}_{2,4}(t)}{2\pi}=\frac{2J_{z}a}{\pi}$ in the half-filling limit $k_{F}=\frac{\pi}{2a}$. Those expressions are valid near the $XY$ phase where $J_{z}\approx 0$ and $K=1$. The phase diagram of (\ref{eq:bosonized_spin_chain}) is described in \cite{Giamarchibook}, where it is demonstrated that the umklapp term is irrelevant for $K>1/2$ and that the system reduces to a TLL.
\section{Derivation of the time-dependent Hamiltonian \label{app:calculations_TDTL}}
This appendix details the derivation of the Hamiltonian (\ref{eq:main_equation}). For ease of notation, we omit the $p$ dependence of $\gamma$ and $\zeta$. The conjugation of $F(I_{0})$ by the  the first unitary transformation $U(\gamma)$  given in Eq. (\ref{eq:U(rho)}) leads to
\begin{eqnarray}
U(\gamma)F(I_{0})U^{\dagger}(\gamma)=\sum_{p\neq 0}\frac{\hbar \omega_{0p}}{\sigma^{2}_{t}}\left\{\cosh[2 \ln(\gamma)]K_{0}(p)-\frac{1}{2}\sinh[2 \ln(\gamma)][K_{+}(p)+K_{-}(p)]\right\}.
\end{eqnarray}
We use the identities with $V(\zeta)$ defined in the main text (\ref{eq:V(alpha)}):
{\small
\begin{eqnarray}
V(\zeta)K_{0}(p)V^{\dagger}(\zeta)&=&K_{0}(p)+\frac{i\zeta}{\omega_{0p}}\left[K_{-}(p)-K_{+}(p)\right]+\left(\frac{\zeta}{\omega_{0p}}\right)^{2}[2K_{0}(p)+K_{+}(p)+K_{-}(p)],\\
V(\zeta)[K_{+}(p)+K_{-}(p)]V^{\dagger}(\zeta)&=&[K_{+}(p)+K_{-}(p)]+\frac{2i\zeta}{\omega_{0p}}[K_{+}(p)-K_{-}(p)]-2\left(\frac{\zeta}{\omega_{0p}}\right)^{2}[K_{+}(p)+K_{-}(p)+2K_{0}(p)],\nonumber\\
\\
V(\zeta)[K_{+}(p)-K_{-}(p)]V^{\dagger}(\zeta)&=&[K_{+}(p)-K_{-}(p)]+\frac{2i\zeta}{\omega_{0p}}[2K_{0}(p)+K_{+}(p)+K_{-}(p)].
\end{eqnarray}}
Furthermore,
\begin{eqnarray}
i\frac{dQ}{dt}Q^{\dagger}=\sum_{p\neq 0}-\left[\frac{\dot{\zeta}}{2\omega_{0p}}+\frac{\zeta}{\omega_{0p}}\frac{\dot{\gamma}}{\gamma}\right][2K_{0}(p)+K_{+}(p)+K_{-}(p)]+i\sum_{p\neq 0} \frac{\dot{\gamma}}{2\gamma}[K_{+}(p)-K_{-}(p)].
\end{eqnarray}
After some algebra, one obtains the expression for $\mathcal{H}(t)$ given by Eq. (\ref{eq:main_equation}).
\section{Nonequilibrium Mean Energy \label{eq:mean_energy}}
The nonadiabatic mean energy in the driven TLL reads
\begin{eqnarray}
\langle \Psi(t)|\mathcal{H}(t)|\Psi(t)\rangle&=&\langle \Psi (0)|Q^{\dagger}\left(Q\frac{I_{0}}{\sigma^{2}_{t}}Q^{\dagger}+i\hbar\frac{\partial Q}{\partial t}Q^{\dagger}\right)Q|\Psi(0)\rangle\nonumber\\
&=&\langle \Psi(0)|\frac{I_{0}}{\sigma^{2}_{t}}|\Psi(0)\rangle+i\hbar\langle\Psi(0)|Q^{\dagger}\frac{\partial Q}{\partial t}|\Psi(0)\rangle.
\end{eqnarray}
Consider the identities
\begin{eqnarray}
U^{\dagger}(\gamma)(K_{+}+K_{-})U(\gamma)&=&\sum_{p\neq 0}\cosh[2\ln(\gamma)](K_{+}+K_{-})+\sum_{p\neq0}2\sinh[2\ln(\gamma)]K_{0},\\
U^{\dagger}(\gamma)K_{0}U(\gamma)&=&\sum_{p\neq 0}\cosh[2\ln(\gamma)]K_{0}+\sum_{p\neq0}\frac{1}{2}\sinh[2\ln(\gamma)](K_{+}+K_{-}),
\end{eqnarray}
where we have eased the notation by keeping the momentum dependence of $\{K_{0}(p),K_\pm(p)\}$ implicit.  
Using these identities, one finds that 
\begin{eqnarray}
i\hbar Q^{\dagger}\frac{\partial Q}{\partial t}&=&-\sum_{p\neq0}\frac{\hbar}{\omega_{0p}}\dot{\zeta}\gamma^{2}\left[K_{0}+\frac{1}{2}(K_{+}+K_{-})\right]+i\hbar \sum_{p\neq0}\frac{\dot{\gamma}}{2\gamma}(K_{+}-K_{-}).
\end{eqnarray}
As a result, the explicit evaluation of the nonadiabatic mean energy yields 
{\small
\begin{eqnarray}
\langle \Psi(t)|\mathcal{H}(t)|\Psi (t)\rangle&=&\langle \Psi(0)|\frac{I_{0}}{\sigma^{2}_{t}}|\Psi(0)\rangle-\sum_{p\neq0}\frac{\hbar}{2\omega_{0p}}\dot{\zeta}\gamma^{2}\langle\Psi(0)|2K_{0}+K_{+}+K_{-}|\Psi(0)\rangle+\sum_{p\neq0}i\hbar \frac{\dot{\gamma}}{2\gamma}\langle \Psi (0)| K_{+}-K_{-}|\Psi(0)\rangle.
\end{eqnarray}}
In particular, for $\zeta=\frac{1}{2}\dot{\gamma}/\gamma$ and $\sigma^{2}_{t}=\gamma^{2}$, 
{\small
\begin{eqnarray}
\langle \Psi(t)|\mathcal{H}(t)|\Psi(t)\rangle&=&\frac{\langle \Psi(0)| I_{0}|\Psi (0)\rangle}{\gamma^{2}}-\sum_{p\neq0}\frac{\hbar }{4\omega_{0p}}(\ddot{\gamma}\gamma-\dot{\gamma}^{2})\langle \Psi(0)| 2 K_{0}+K_{+}+K_{-}|\Psi(0)\rangle+\sum_{p\neq0}i\hbar \frac{\dot{\gamma}}{2\gamma}\langle \Psi(0)|K_{+}-K_{-}|\Psi(0)\rangle.\nonumber\\
\end{eqnarray}}
Assuming an initial thermal state $\rho_{0}$ and denoting the expectation value  by $\langle \bullet \rangle_{0}\equiv {\rm tr}(\rho_{0} \bullet)$, it follows that
\begin{eqnarray}
\langle \mathcal{H}(t)\rangle&=&\sum_{p\neq 0}\hbar\omega_{0p}\frac{\langle K_{0}\rangle_{0}}{\gamma^{2}}-\sum_{p\neq 0}\frac{\hbar }{4\omega_{0p}}(\ddot{\gamma}\gamma-\dot{\gamma}^{2}) \langle 2 K_{0}+K_{+}+K_{-}\rangle_{0}+\sum_{p\neq0}i\hbar \frac{\dot{\gamma}}{2\gamma}\langle K_{+}-K_{-}\rangle_{0}\nonumber\\
\\
&=&\sum_{p\neq 0}\left[\frac{\hbar\omega_{0p}}{\gamma^{2}}-\frac{\hbar }{2\omega_{0p}}(\ddot{\gamma}\gamma-\dot{\gamma}^{2}) \right]\left[\langle n_{B}(p)\rangle +\frac{1}{2}\right]+\sum_{p\neq0}i\hbar \frac{\dot{\gamma}}{2\gamma}\langle K_{+}-K_{-}\rangle_{0}\nonumber\\
&=&\sum_{p \neq 0}\langle \mathcal{H}(t)\rangle_{p},
\end{eqnarray}
with the Bose distribution $\langle n_{B}(p)\rangle =1/[\exp(2\beta_{0}\hbar\omega_{0p})-1]$. Making use of the Ermakov equation $\ddot{\gamma}=-\Omega^{2}(p,t)\gamma+\frac{\omega^{2}_{0p}}{\gamma^{3}}$, one can also express the mean energy as
\begin{eqnarray}
\langle \mathcal{H}(t)\rangle&=&\sum_{p\neq 0}\left\{\frac{\hbar\omega_{0p}}{2\gamma^{2}}+\frac{\hbar }{2\omega_{0p}}[\Omega^{2}(p,t)\gamma^{2}+\dot{\gamma}^{2}]\right\} \left[\langle n_{B}(p)\rangle +\frac{1}{2}\right]+\sum_{p\neq0}i\hbar \frac{\dot{\gamma}}{2\gamma}\langle K_{+}-K_{-}\rangle_{0}\nonumber\\
&=&\sum_{p \neq 0}\langle \mathcal{H}(t)\rangle_{p}.
\end{eqnarray}
This recovers the expression of the mean energy established in \cite{Lewis1969,Beau20}. 
\subsection{Adiabatic mean energy}
It is interesting to consider the limit of an adiabatic quench. In the adiabatic limit $\dot{\gamma}=\ddot{\gamma}=0$ and $\gamma=\sqrt{\omega_{0p}/\Omega(p,t)}$, one recovers the Hamiltonian diagonalized by the Bogoliubov transformation in the case $g_{2}(p,t)=g_{4}(p,t)$, that is, 
\begin{align}
\langle \mathcal{H}(t)\rangle&=\sum_{p\neq 0}\hbar \Omega(p,t)\left[\langle n_{B}(p)\rangle +\frac{1}{2}\right]\\
&=\sum_{p\neq 0}\hbar \sqrt{\omega_{0p}(\omega_{0p}+2g(p,t))}\left[\langle n_{B}(p)\rangle +\frac{1}{2}\right]\\
&=\sum_{p\neq 0}\hbar \epsilon(p,t)\left[\langle n_{B}(p)\rangle +\frac{1}{2}\right].
\end{align}
For an adiabatic quench, the final state is the adiabatic state, and the final energy should be given by the expectation value of the Hamiltonian diagonalized by the Bogoliubov transformation. Thus,   for a pure state,
\begin{align}
\langle \mathcal{H}(\tau_{Q}\to \infty)\rangle&=\frac{1}{2}\sum_{p\neq 0}[\epsilon(p,\tau_{Q}\to \infty)]\nonumber\\
&=\frac{L}{2\pi}\int_{0}^{\infty}dp\ \hbar p\sqrt{v^{2}_{F}+2v_{F}\alpha}e^{-R_{0}p}\nonumber\\
&=\frac{L}{2\pi}\frac{\hbar}{R^{2}_{0}}\sqrt{v_{F}(v_{F}+2\alpha)}\nonumber\\
&\approx \frac{L}{2\pi}\frac{\hbar}{R^{2}_{0}}\left(v_{F}+\alpha-\frac{1}{2}\frac{\alpha^{2}}{v_{F}}\right).
\end{align}
Note that the very last Taylor expansion is only valid for  $2\alpha \ll v_{F}$. 
\subsection{Mean energy following a sudden quench}
For a consistency check, one can compute the sudden quench energy $\tau_{Q}\to 0$. Using the fact that the state immediately after the quench equals the initial ground state, only the $K_{0}(p)$ operator of the Hamiltonian contributes, giving
\begin{eqnarray}
\langle \mathcal{H}(\tau_{Q}\to 0)\rangle\frac{2\pi}{L}=2\int_{0}^{\infty}\hbar \omega(p,t)\frac{1}{2}e^{-R_{0}p}dp=\frac{\hbar(v_{F}+\alpha)}{R^{2}_{0}}.
\end{eqnarray}
As a consequence, the non-adiabatic residual energy in a sudden quench is given by
\begin{align}
\Delta Q(\tau_{Q} \to 0)&=\langle \mathcal{H}(\tau_{Q}\to 0)\rangle-\langle \mathcal{H}(\tau_{Q}\to \infty)\rangle\nonumber\\
&=\frac{L}{2\pi}\frac{\hbar \alpha^{2}}{2 R^{2}_{0}v_{F}}.
\end{align}

\subsection{Mean energy for the inverse polynomial quench}
We now discuss the mean energy for the driving protocol of section \ref{section_polynomial_quench}. If one takes $\gamma=Bt+1$ to ensure the driving (\ref{eq:lambda}), the mean energy reduces to 
\begin{eqnarray}
\langle \mathcal{H}(t)\rangle_{p}=\hbar\omega_{0p}\left[\langle n_{B}(p)\rangle +\frac{1}{2}\right]\frac{1}{(Bt+1)^{2}}+\frac{\hbar B^{2} }{2\omega_{0p}} \left[\langle n_{B}(p)\rangle +\frac{1}{2}\right]+i\hbar\frac{B}{2(Bt+1)}\langle K_{+}-K_{-}\rangle_{0}.
\end{eqnarray}
In the case of the pure ground state,  corresponding to $T\to 0$ and $\langle n_{B}(p)\rangle=0$, it follows that 
\begin{eqnarray}
\langle \mathcal{H}(t)\rangle_{p}=\frac{\hbar\omega_{0p}}{2}\frac{1}{(Bt+1)^{2}}+\frac{\hbar B^{2}}{4\omega_{0p}}.
\end{eqnarray}
The residual excitation at the final time is 
\begin{align}
\langle \mathcal{H}(\tau_{Q})\rangle&=\frac{1}{2}\frac{1}{(B\tau_{Q}+1)^{2}}\sum_{p\neq 0}\hbar\omega_{0p}+\frac{\hbar B^{2}}{4}\sum_{p\neq0}\frac{1}{\omega_{0p}}\nonumber\\
&=\frac{1}{(B\tau_{Q}+1)^{2}}\frac{L}{2\pi}\int_{\frac{2\pi}{L}}^{\infty} dp e^{-R_{0}p}\hbar\omega_{0p}+\frac{\hbar B^{2}}{2}\frac{L}{2\pi}\int_{\frac{2\pi}{L}}^{\infty} dp e^{-R_{0}p}\frac{1}{\omega_{0p}}\nonumber\\
&=\frac{1}{(B\tau_{Q}+1)^{2}}\frac{L}{2\pi}\hbar v_{F}\frac{e^{-\frac{2\pi}{L}R_{0}}(1+\frac{2\pi}{L}R_{0})}{R^{2}_{0}}+\frac{\hbar B^{2}}{2}\frac{L}{2\pi}\frac{1}{v_{F}}\Gamma(0,\frac{2\pi}{L}R_{0}),\nonumber\\
\end{align}
where the incomplete gamma function is defined as $\Gamma(s,x)=\int_{x}^{\infty}t^{s-1}e^{-t}dt$. In the adiabatic limit $B\to 0$ and, 
\begin{align}
\langle \mathcal{H}(\tau_{Q}\to\infty)\rangle&=\left(\frac{\tilde{g}_{2,4}(\tau_{Q})}{\pi v_{F}}+1\right)^{1/2}\frac{L}{2\pi}\hbar v_{F}\frac{e^{-\frac{2\pi}{L}R_{0}}(1+\frac{2\pi}{L}R_{0})}{R^{2}_{0}}.
\end{align}
Also, in the thermodynamic limit $L\to \infty$, one recovers the expected adiabatic limit obtained from the Bogoliubov transformation
\begin{align}
\langle \mathcal{H}(\tau_{Q}\to\infty)\rangle&=\frac{\hbar}{R^{2}_{0}}\frac{L}{2\pi}\sqrt{v_{F}\left(v_{F}+2\frac{\tilde{g}_{2,4}(\tau_{Q})}{2\pi}\right)}.
\end{align}
As a consequence, one can express the non-adiabatic residual energy as
\begin{align}
\Delta Q(\tau_{Q})&=\frac{\hbar B^{2}}{2}\frac{L}{2\pi}\frac{1}{v_{F}}\Gamma(0,\frac{2\pi}{L}R_{0}).
\end{align}
In particular, in the perturbative regime $\frac{\tilde{g}_{2,4}(\tau_{Q})}{\pi v_{F}} \ll 1$,
$B=\frac{1}{\tau_{Q}}\left[-1\pm\left(\frac{\tilde{g}_{2,4}(\tau_{Q})}{\pi v_{F}}+1\right)^{-1/4}\right]\approx -\frac{1}{\tau_{Q}}\frac{\tilde{g}_{2,4}(\tau_{Q})}{4\pi v_{F}}$. Hence, the non-adiabatic residual energy scales as
\begin{align}
\Delta Q(\tau_{Q})&=\frac{1}{\tau^{2}_{Q}}\frac{\hbar L}{2\pi}\left(\frac{\tilde{g}_{2,4}(\tau_{Q})}{2\pi}\right)^{2}\frac{1}{4v^{2}_{F}}\frac{1}{2v_{F}}\Gamma(0,\frac{2\pi}{L}R_{0}),\\
\end{align}
leading to the expression in the main text Eq. (\ref{eq:perturbative_reverse_engineering}).
\section{Details on the calculations for the STA assisted by a semi-classical sine-Gordon potential}
In this Appendix, we provide extra details on the computation of the semiclassical sine-Gordon Hamiltonian (\ref{eq:sine_gordon}). We start by demonstrating that the choice $\zeta_{p}=\frac{\dot{\gamma}_{p}\sigma^{2}_{t}}{2\gamma^{3}_{p}}$ in Eq. (\ref{eq:main_equation}), where $\gamma_{p}=\gamma$ and $\sigma_{t}$ are scalars independent of the momentum $p$, recovers the semiclassical approximation of the sine-Gordon model. This corresponds to the most general case $g_{2}(p,t)\neq g_{4}(p,t)$. In the momentum representation, after insertion of the parameters in (\ref{eq:main_equation})
\small
\begin{align}
\mathcal{H}(t)&=\sum_{p\neq 0}\frac{\hbar \omega_{0p}}{2}\frac{1}{(\sigma_{t}\gamma)^{2}}[K_{0}(p)+\frac{1}{2}(K_{+}(p)+K_{-}(p))]\nonumber\\
&+\sum_{p\neq 0}\frac{\hbar \omega_{0p}}{2}\left(\frac{\gamma}{\sigma_{t}}\right)^{2}[K_{0}(p)-\frac{1}{2}(K_{+}(p)+K_{-}(p))]\nonumber\\
&+\sum_{p\neq 0}\frac{\hbar}{\omega_{0p}}[K_{0}(p)+\frac{1}{2}(K_{+}(p)+K_{-}(p))]\nonumber\\
&\times\left(\frac{\sigma_{t}}{\gamma}\right)^{2}\left\{ \left(\frac{\dot{\gamma}}{\gamma}\right)^{2}-\frac{1}{2}\frac{\ddot{\gamma}}{\gamma}-\frac{\dot{\gamma}}{\gamma}\frac{\dot{\sigma}_{t}}{\sigma_{t}}\right\}.
\end{align}\normalsize
Note that the last term of the Hamiltonian has, as it should,  dimensions of energy, given that $[\dot{\gamma}]=[\dot{\sigma}_{t}]=[T^{-1}]$,  $[\ddot{\gamma}]=[T^{-2}]$ and $[\omega_{0p}]=[T^{-1}]$, where $[\bullet]$ stands for the dimension. Using the relations between the bosonic and the field basis, it is now possible to express Eq. (\ref{eq:sine_gordon}).
\section{Adding an external potential}
Inserting a periodic external potential potential $V(x,t)=V_{0}(t)\cos(\lambda_{0} x+\phi_{0})$ yields 
\begin{eqnarray}
H_{V}&=&\int_{0}^{L}dx\rho(x)V(x)\approx V_{0}\int_{0}^{L}dx\cos(\lambda_{0} x+\phi_{0})\rho_{0}\left\{1+2\cos[2(\pi\rho_{0}x-\theta(x))]\right\}\\
&=&V_{0}\int_{0}^{L}dx\rho_{0}\left\{\cos(\lambda_{0} x+\phi_{0})+\cos[2(\pi\rho_{0}x-\theta(x))+\lambda_{0} x]+\cos[2(\pi\rho_{0}x-\theta(x))-\lambda_{0} x]\right\}.
\end{eqnarray}
Choosing the periodicity of the lattice commensurate to the density $\lambda_{0}=2\pi\rho_{0}$ and $\phi_{0}=\pi$ leads to 
\begin{eqnarray}
H_{V}&=&V_{0}(t)\int_{0}^{L}dx\rho_{0}\left\{\cos(2\pi\rho_{0} x+\pi)+\cos[4\pi\rho_{0}x-2\theta(x)+\pi]+\cos[-2\theta(x)+\pi]\right\}\\
&\approx&-V_{0}(t)\rho_{0}\int_{0}^{L}dx\cos[2\theta(x)].
\end{eqnarray}
The first cosine vanishes upon integration as $\rho_{0}=N/L$, and the second drops considering that $\theta(x)$ varies slowly compared to the strongly oscillatory part $4\pi\rho_{0}x$. For a large enough potential $V_{0}(t)$, the field $\theta(x)$ is locked near zero so that the cosine minimizes the energy of the Hamiltonian. One can then Taylor expand for small fluctuations
\begin{eqnarray}
H_{V}&\approx& 2V_{0}(t)\rho_{0}\int_{0}^{L}dx[\theta(x)]^{2}-V_{0}(t)\rho_{0}L.
\end{eqnarray}
As a result, one obtains the semiclassical limit of the sine-Gordon Hamiltonian given in the main text (\ref{eq:sg_model}).
\section{General solution to the Ermakov-Pinney equation}\label{AppPinney}
This section details the Pinney solution of the Ermakov equation \cite{Pinney50}. The Pinney solution of  Eq. (\ref{eq:Ermakov}) is expressed in the form
\begin{eqnarray}
\gamma_{p}(t)&=&\left[u^{2}(t)+\frac{\omega^{2}_{0p}}{(W[u,v])^{2}}v^{2}(t)\right]^{1/2},
\end{eqnarray}
where $u$ and $v$ are solutions of the homogeneous equation $\ddot{\gamma}_{p}+\Omega^{2}(p,t)\gamma_{p}=0$ with initial conditions $u(0)=\gamma_{0},\dot{u}(0)=\dot{\gamma}_{0},v(0)=0,\dot{v}(0)\neq0$. The Wronskian is defined as $W[u,v]=u\dot{v}-v\dot{u}$. The observation that the Wronskian is a constant follows from its vanishing derivative $\dot{W}[u,v]=u\ddot{v}-v\ddot{u}=0$. We want to determine the most general form for the functions $u$ and $v$ that satisfy the given initial conditions. Let us consider the form 
\begin{eqnarray}
u(t)&=&c_{1}r(t)+c_{2}s(t),\\
v(t)&=&c'_{1}r(t)+c'_{2}s(t).
\end{eqnarray}
The coefficients $c_{1}$ and $c_{2}$ can be expressed as
\begin{eqnarray}
c_{1}&=&\frac{W[u,s]}{W[r,s]},\\
c_{2}&=&\frac{W[u,r]}{W[s,r]}.
\end{eqnarray}
Using the initial conditions yields 
\begin{align*}
c_{2}=\frac{\gamma_{0}\dot{r}(0)-\dot{\gamma}_{0}r(0)}{W[s,r]},&\quad c_{1}=\frac{\gamma_{0}\dot{s}(0)-\dot{\gamma}_{0}s(0)}{W[r,s]},\\
c'_{2}=\frac{-\dot{v}(0)r(0)}{W[s,r]},&\quad c'_{1}=-\frac{\dot{v}(0)s(0)}{W[r,s]}.
\end{align*}
In order to simplify the Wronskian $W[u,v]$, a convenient choice is $\dot{v}(0)=\frac{1}{\gamma_{0}}$, as it leads to  $W[u,v]=u(0)\dot{v}(0)=1$. It follows that
\begin{align*}
c_{2}=\frac{\gamma_{0}\dot{r}(0)-\dot{\gamma}_{0}r(0)}{W[s,r]},&\quad c_{1}=\frac{\gamma_{0}\dot{s}(0)-\dot{\gamma}_{0}s(0)}{W[r,s]},\\
c'_{2}=\frac{-r(0)}{\gamma_{0}W[s,r]},&\quad c'_{1}=-\frac{s(0)}{\gamma_{0}W[r,s]}.
\end{align*}
We also note that the choice $\dot{\gamma}_{0}=0$ leads to further simplifications
\begin{eqnarray}
c_{2}=\frac{\gamma_{0}\dot{r}(0)}{W[s,r]},\quad c_{1}=-\frac{\gamma_{0}\dot{s}(0)}{W[s,r]}.
\end{eqnarray}
Finally, we obtain the general expressions given in the main text, Eqs.  (\ref{eq:u_general}) and (\ref{eq:v_general}).
\section{Solutions to the Airy Ermakov equation}\label{AppAiry}
We are interested in solving the second-order homogeneous differential equation of the form 
\begin{eqnarray}
\ddot{\gamma}_{p}+(at+b)\gamma_{p}=0. \label{linear_homogeneous_Airy}
\end{eqnarray}
Making the change of variable $z(t)=-\frac{at+b}{(-a)^{2/3}}$, the equation takes the form
\begin{eqnarray}
\frac{\partial^{2}\gamma_{p}}{\partial z^{2}}-z\gamma_{p}=0,
\end{eqnarray}
which admits solutions in terms of the Airy functions \cite{abramowitz}. The linearly independent solutions of equation (\ref{linear_homogeneous_Airy}) are then found in the form
\begin{eqnarray}
r(t)={\rm Ai}\left[z(t)\right],\\
s(t)={\rm Bi}\left[z(t)\right].
\end{eqnarray}
The derivative of the Airy function reads
\begin{eqnarray}
\partial_{t}{\rm Ai}[z]&=&(-a)^{1/3}\partial_{z}{\rm Ai}[z],\\
\partial_{t}{\rm Bi}[z]&=&(-a)^{1/3}\partial_{z}{\rm Bi}[z],
\end{eqnarray}
while the Wronskian reduces to 
\begin{eqnarray}
W_{t}[{\rm Ai}(z),{\rm Bi}(z)]&=&\frac{\partial z}{\partial t}W_{z}[{\rm Ai}(z),{\rm Bi}(z)]=(-a)^{1/3}\pi^{-1}.
\end{eqnarray}
Finally, we obtain the system of equations in the main text, Eq. (\ref{eq_Airy_Ermakov}).
\section{Perturbative solution to the Ermakov equation}
In this section, we present a perturbative solution to the Ermakov equation for a small interaction quench $\alpha \ll v_{F}/2$, which controls the amplitude of $a$. Let us consider the Ermakov equation
\begin{align}
\ddot{\gamma}_{p}\gamma^{3}_{p}+(at+b)\gamma^{4}_{p}=\omega^{2}_{0p}.
\end{align}
One can solve this equation to second order with the solution $\gamma_{p}=\gamma^{(0)}+a \gamma^{(1)}+a^{2}\gamma^{(2)}$, where the zeroth, first and second order parameters verify 
\begin{align}
\ddot{\gamma}^{(0)}+b\gamma^{(0)}&=\frac{\omega^{2}_{0p}}{{\gamma^{(0)}}^{3}},\\
\ddot{\gamma}^{(1)}+\left[\frac{3\omega^{2}_{0p}}{{\gamma^{(0)}}^{4}}+b\right]\gamma^{(1)}&=-t\gamma^{(0)},\\
\ddot{\gamma}^{(2)}+\left[\frac{3\omega^{2}_{0p}}{{\gamma^{(0)}}^{4}}+b\right]\gamma^{(2)}&=-t\gamma^{(1)}+6\frac{\omega^{2}_{0p}}{{\gamma^{(0)}}^{5}}{\gamma^{(1)}}^{2}.
\end{align}
Furthermore, choosing $b=\omega^{2}_{0p}$, the solution to the zeroth order equation with $\dot{\gamma}_{0}=0$ reads
\begin{align}
\gamma^{(0)}=\sqrt{\gamma_{0}^{2}+\left[\left(\frac{1}{\gamma_{0}}\right)^{2}-\gamma^{2}_{0}\right]\sin^{2}(\omega_{0p}t)}.
\end{align}
In particular, for $\gamma_{0}=1$, one obtains
\begin{align}
\gamma^{(0)}=1.
\end{align}
The differential equation for the first-order solution is 
\begin{align}
\ddot{\gamma}^{(1)}+4\omega^{2}_{0p}\gamma^{(1)}&=-t,
\end{align}
that is solved by
\begin{align}
\gamma^{(1)}=\frac{[\sin(2\omega_{0p}t)-2t\omega_{0p}]}{8\omega^{3}_{0p}}.
\end{align}
Similarly, for the second order, 
\begin{align}
\ddot{\gamma}^{(2)}+4\omega^{2}_{0p}\gamma^{(2)}&=-t\gamma^{(1)}+6\omega^{2}_{0p}\gamma^{(1)2},
\end{align}
the solution reads
\begin{align}
    \gamma^{(2)}=\frac{-\{[9+\cos(2t\omega_{0p})]\sin^{2}(t\omega_{0p})\}+2t\omega_{0p}\{-\sin(2t\omega_{0p})+t[5+2\cos(2t\omega_{0p})]\omega_{0p}\}}{64 \omega^{6}_{0p}}.
\end{align}
\section{Perturbative mean energy \label{perturbative_airy}}
It would be interesting to determine a perturbative result for small quenches for the mean energy. One can use the previously obtained expression $\gamma=\gamma^{(0)}+a\gamma^{(1)}+a^{2}\gamma^{(2)}$ in the equation for the nonadiabatic mean energy
\begin{eqnarray}
\langle \mathcal{H}(t)\rangle&=&\sum_{p\neq 0}\left[\frac{\hbar\omega_{0p}}{\gamma^{2}}-\frac{\hbar }{2\omega_{0p}}(\ddot{\gamma}\gamma-\dot{\gamma}^{2}) \right]\left[\langle n_{B}(p)\rangle +\frac{1}{2}\right]+\sum_{p\neq0}i\hbar \frac{\dot{\gamma}}{2\gamma}\langle K_{+}-K_{-}\rangle_{0}\nonumber\\
&=&\sum_{p\neq 0}\frac{\hbar}{2\omega_{0p}}\left[\ddot{\gamma}\gamma+2\Omega^{2}(p,t)\gamma^{2}+\dot{\gamma}^{2} \right]\left[\langle n_{B}(p)\rangle +\frac{1}{2}\right]+\sum_{p\neq0}i\hbar \frac{\dot{\gamma}}{2\gamma}\langle K_{+}-K_{-}\rangle_{0}.
\end{eqnarray}
In order to compute the sum, we assume that the initial state is a pure state $\langle n_{B}(p)\rangle=0$. In the thermodynamic limit, one can express the mean energy as
\begin{align}
\langle \mathcal{H}(t)\rangle\frac{2\pi}{L}&=\int_{0}^{\infty}dp \frac{\hbar}{2\omega_{0p}}\left[\ddot{\gamma}\gamma+2\Omega^{2}(p,t)\gamma^{2}+\dot{\gamma}^{2} \right]e^{-R_{0}p}.\label{eq:mean_energy_integral}\\
\end{align}
Inserting the perturbative solution in the mean energy in the above expression (\ref{eq:mean_energy_integral}) leads to the expression evaluated at the final quench time $\tau_{Q}$
\begin{align}
    \langle \mathcal{H}(\tau_{Q})\rangle\frac{2\pi}{L}&=\int_{0}^{\infty}dp \hbar \left[p(v_{F}+\alpha-\frac{\alpha^{2}}{2v_{F}})+\frac{\alpha^{2}\sin^{2}(v_{F}p\tau_{Q})}{2pv^{3}_{F}\tau^{2}_{Q}}\right]e^{-R_{0}p}\\
    &=\frac{v_{F}+\alpha}{R^{2}_{0}}-\frac{\alpha^{2}}{2v_{F}R^{2}_{0}}+E_{gs}\left(\frac{\tau_{0}}{\tau_{Q}}\right)^{2}\ln\left(1+\left(\frac{\tau_{0}}{\tau_{Q}}\right)^{2}\right),
\end{align}
with $E_{gs}=\frac{\hbar L}{2\pi}\frac{\alpha^{2}}{2v_{F}R^{2}_{0}}$ and $\tau_{0}=R_{0}/(2v_{F})$. It follows that the non-adiabatic residual energy is given by Eq. (\ref{eq:deltaQ_Ermakov}) in the main text.

\twocolumngrid

\bibliography{Luttinger}
\newpage
\end{document}